\documentclass[preprint,12pt]{elsarticle}

\journal{Icarus}

\usepackage{natbib}
\usepackage{aas_macros,amsmath}
\usepackage{url}
\usepackage{array}
\usepackage{textcomp}
\usepackage{amssymb}
\usepackage{siunitx}

\usepackage{todonotes}

\newcommand{\rev}[1]{{#1}}
\newcommand{\revrev}[1]{{#1}}
\newcommand{\revrevrev}[1]{{#1}}

\biboptions{authoryear}

\begin{document}
\begin{frontmatter}

\title{Exogenous delivery of water to Mercury}

\author[kapteynaddress,sronaddress]{Kateryna~Frantseva\corref{mycorrespondingauthor}}
\cortext[mycorrespondingauthor]{Corresponding author}
\ead{frantseva@astro.rug.nl}

\author[usaddress]{David~Nesvorn{\'y}}

\author[novaaddress,leidenaddress,sronaddress]{Michael~Mueller}

\author[sronaddress,kapteynaddress]{Floris~F.S.~van~der~Tak}

\author[utrechtaddress]{Inge~Loes~ten~Kate}

\author[petraddress,nasaaddress,cresstaddress]{Petr~Pokorn{\'y}}

\address[kapteynaddress]{Kapteyn Astronomical Institute, University of Groningen, Landleven 12, 9747 AD Groningen, The Netherlands}
\address[sronaddress]{SRON Netherlands Institute for Space Research, Landleven 12, 9747 AD Groningen, The Netherlands}
\address[usaddress]{Department of Space Studies, Southwest Research Institute, 1050 Walnut Street, Boulder, CO 80302, USA}
\address[novaaddress]{NOVA Netherlands Research School for Astronomy, The Netherlands}
\address[leidenaddress]{Leiden Observatory, University of Leiden, Niels Bohrweg 2, 2333 CA Leiden, The Netherlands}
\address[utrechtaddress]{Department of Earth Sciences, Utrecht University, Budapestlaan 4, 3584 CD Utrecht, The Netherlands}
\address[petraddress]{Department of Physics, The Catholic University of America, 620 Michigan Ave, NE Washington, DC 20064, USA}
\address[nasaaddress]{Astrophysics Science Division, NASA Goddard Space Flight Center, 8800 Greenbelt Rd., Greenbelt, MD 20771, USA}
\address[cresstaddress]{Center for Research and Exploration in Space Science and Technology, NASA/GSFC, Greenbelt, MD 20771, USA}

\begin{abstract}
 
Radar and \revrev{spacecraft} observations \rev{show the permanently shadowed regions around Mercury's North Pole to} contain water \rev{ice} and complex organic material. \rev{One possible source of this material are impacts by} interplanetary dust particles (IDPs), asteroids, and comets.

We have performed numerical simulations of the \revrev{dynamical evolution} of asteroids and comets over the few Myr and \revrev{checked for their impacts with Mercury}. We use the N-body integrator RMVS/Swifter to propagate the Sun and the eight planets from their current positions. We add comets and asteroids to the simulations as massless test particles, based on their current orbital distributions. Asteroid impactors are assigned a probability of being water-rich (C-class) based on the measured distribution of taxonomic types. For comets, we assume a constant water fraction. For IDPs, we use a dynamical \revrev{meteoroid model} to compute the dust flux on Mercury. \rev{Relative to previous work on asteroid and comet impacts \citep{Moses1999}, we leverage 20 years of progress in minor body surveys.}

Immediate post-impact ejection of impactor material into outer space is taken into account as is the migration efficiency of water across Mercury's surface to the polar cold traps.

We find that asteroids deliver \rev{$\sim 1 \times 10^{3}$\,kg/yr of water to Mercury, comets deliver $\sim 1 \times 10^{3}$\,kg/yr and IDPs deliver $\sim 16 \times 10^{3}$\,kg/yr within a factor of several. Over a timescale of $\sim 1$\,Gyr, this is enough to deliver the minimum amount of water required by the radar and MESSENGER observations.}  
 
\rev{While other sources of water on Mercury are not ruled out by our analysis, we show that they are not required to explain the \revrev{currently available observational lower limits.}}

\end{abstract}


\end{frontmatter}

%\linenumbers

\section{Introduction}\label{introduction}

The \rev{presence} of water in the permanently shadowed polar regions of Mercury is well established. \rev{Through} ground-based observations of Mercury bright radar-reflective regions were detected near the poles of the planet \citep{Slade1992,Harmon1992,Butler1993,Harmon1994} that are consistent with water ice (although sulphur and certain silicates could also explain the radar data). 
Based on the radar observations, \citet{Moses1999} determined the total mass of the ice to be \rev{$4\times 10^{13} - 8\times 10^{14}$\,kg} assuming a layer thickness  between 2\,m and 20\,m.

Observations with the Neutron Spectrometer (NS) aboard the MErcury Surface, Space ENvironment, GEochemistry, and Ranging (MESSENGER) spacecraft \citep{Lawrence2013} demonstrated that the radar-bright deposits in the North polar region \rev{consist of} water ice. NS was not sensitive to the South polar region due to the eccentric spacecraft orbit. Specifically, NS observations identified a hydrogen-rich layer associated with water ice of unknown thickness, covered by a less hydrogen-rich layer (less than 25 weight \% water-equivalent hydrogen) with a thickness between \rev{0.1 and 0.3\,m}. \citet{Lawrence2013} adopted an ice layer thickness between 0.5\,m and 20\,m, where the lower limit follows from the models of \revrev{the} surface modification processes \citep{crider2005} and the upper limit is estimated from models of radar scattering \citep{Butler1993}. 

Additional MESSENGER observations using the Mercury Laser Altimeter (MLA) and the Mercury Dual Imaging System (MDIS) allowed both polar regions to be mapped three-dimensionally, and the area of the permanently shadowed regions to be measured: \rev{$(1.25 - 1.46) \times 10^{10}$ m$^2$} around the north pole and \rev{$(4.3 \pm 1.4) \times 10^{10}$ m$^{2}$} around the south pole \citep{chabot2012,Neumann2013}.

Combining these two sets of MESSENGER results, and assuming that radar-reflective regions in Mercury's South polar region are also dominantly water ice with the same layer thickness as in the North, the MESSENGER observations imply a total water mass at Mercury's poles of \rev{$2.1 \times 10^{13}$ to $1.4 \times 10^{15}$\,kg} \citep{Lawrence2013}, consistent with the interpretation of ground-based radar observations by \citeauthor{Moses1999}. \rev{Independently, \citet{Eke2017} derive an upper limit of $\sim 3\times 10^{15}$\,kg from MLA observations of \revrev{the} craters.}

\revrev{Various}\rev{ ideas have been presented as to the origin of Mercury's water ice.}
\rev{Endogenous sources such as volcanic activity, crust and mantle outgassing, \revrev{and} interaction of surface rocks with \revrev{the} solar wind might have played a role in the formation of the bright and dark deposits on the surface of Mercury \citep{Potter1995}. \citet{Nittler2017} suggest that Mercury's polar deposits include some fraction of material derived from volatiles outgassed from Mercury's interior, potentially caused by volcanic resurfacing  \citep{Wilson2008,Head2009,Prockter2010,Denevi2013,Ostrach2015}.}

\rev{On the other hand, water has been delivered from outside of Mercury.  In the current Solar System, plausible impactor populations with meaningful water content would be both macroscopic bodies (asteroids and comets) and microscopic impactors (Interplanetary Dust Particles, IDPs). \citet{Frantseva2018} showed that such impacts may be the source of organic material found on the surface of Mars. Could Mercury's water ice have been deposited in a similar fashion?}

\rev{Water ice can have accumulated in Mercury's polar cold traps since the end of the Late Heavy Bombardment $\sim 3.5$\,Gyr ago \citep{Moses1999}} or, equivalently for our purposes, the end of the formation of the terrestrial planets according to the accretion tail scenario \citep{Morbidelli2018}. This idea is reinforced by thermal modeling, demonstrating that the permanently shadowed regions are cold enough to keep water ice thermally stable for billions of years \citep{Paige2013}. Water ice deposited on other parts of Mercury's surface is unstable against sublimation; water vapor will diffuse across the surface until it either re-condensates at a polar cold trap or is dissociated by solar ultraviolet radiation and lost to space. \rev{Between $\sim 5-15$\% of the water deposited across Mercury's surface will reach the polar cold traps \citep{Butler1993,Butler1997};} \rev {observational evidence for this diffusion process occurring on the surface of the Moon has recently been presented by \citet{Hendrix2019}.}

Water delivery by impacts of IDPs, asteroids, and comets has been studied in the past, but questions remain. \citet{Moses1999} estimated the water flux due to impacts of asteroids and comets in the last 3.5\,Gyr using a Monte Carlo simulation to generate fictitious comets and asteroids, basing themselves on the populations known at the time and (large) correction factors accounting for observational incompleteness. They find that impacts from Jupiter-family comets can supply \rev{$(0.1 - 200)\times 10^{13}$\,kg} of water to Mercury's polar regions (corresponding to ice deposits $0.05 - 60$\,m thick), Halley-type, i.e., long-period comets can supply \rev{$(0.2 - 20)\times 10^{13}$\,kg} of water to the poles ($0.07 - 7$\,m of ice), and asteroids can provide \rev{$(0.4 - 20)\times 10^{13}$\,kg} of water to the poles ($0.1 - 8$\,m of ice). In order to calculate the IDP \rev{($10-500\,\mu$m,)} flux on Mercury they extrapolated the current terrestrial influx of IDPs to that at Mercury. The continuous IDP flux on Mercury within the last 3.5\,Gyr would deliver \rev{$(3 - 60)\times 10^{13}$\,kg} of water ice to the permanently shaded regions at Mercury's poles (equivalent to an average ice thickness of $0.8 - 20$\,m). More recently, for particles in the size range of $5-100\,\mu$m, using a numerical code that takes into account the gravitational interaction with all planets and non-gravitational forces such as the Poynting-Robertson drag and the solar wind drag, \citet{Borin2017} have estimated the \rev{total} impact flux to be $1.97\times 10^8$\,kg/yr, 20 times higher than the estimate by \citeauthor{Moses1999} and 5 times larger than the flux on Earth \citep{Love1993}. \rev{Another study of the dust environment around Mercury has been done by \citet{Pokorny2018} for particles in the size range of $10-2000\,\mu$m. The total dust flux following from the study is $4.4\times 10^6$\,kg/yr, similar to \citeauthor{Moses1999} and two orders of magnitude less than \citet{Borin2017}.} 

Given the large discrepancy between the existing estimates, and their large uncertainty due to observational incompleteness, we aim to update the asteroid, comet\rev{,} and dust \rev{impact} rates on Mercury based on the most recent catalogs of minor bodies, leveraging the results of $\sim$20 years of search programs for asteroids and comets \rev{since} \citet{Moses1999}. First, we study the rates at which comets and asteroids impact Mercury in the current Solar System and we derive the corresponding water delivery rates. We do so using $N$-body codes modeling the \rev{motions} of asteroids and comets under the gravitational influence of the Sun and the planets while checking for impacts. \revrev{These models are described in Section\ \ref{method} together with the intrinsic collision probabilities of our asteroid and comet populations and the derived impact rates. Also, we use the ZoDy model from \citet{Nesvorny2010,Nesvorny2011a,Nesvorny2011b} to calculate IDP accretion on Mercury in Section\ \ref{method_idp}. In Section\ \ref{water_delivery}, we calculate the corresponding water delivery rates on Mercury's surface.} As discussed above, a fraction of the water will be lost to space immediately after impact, as well as during migration to the poles; \rev{this fraction is estimated} in Section\ \ref{survivability}. The implications of our findings are discussed in Section\ \ref{discussion}. 

\section{Gravitational dynamics of asteroids and comets}\label{method}

To study Mercury impact rates in geologically recent times we performed numerical simulations of the gravitational dynamics of the current Solar System. Our model accounts for the gravity of the Sun plus the eight planets Mercury to Neptune. Asteroids and comets are added as different sets of passive test particles, i.e., their gravity is neglected. Non-gravitational forces are also neglected. We integrate forward in time over Myr timescales. Over these timescales, the current Solar System is close to a steady state, even when accounting for the non-gravitational Yarkovsky effect, so we can assume impact rates to be constant with time (see \revrev{Section~\ref{method_asteroids}} for a cross-check on this assumption).

\rev{To model the gravitational dynamics, we use the $N$-body integrator RMVS that corresponds to the RMVS3 version of Swift \citep[Regularized Mixed Variable Symplectic;][]{Levison1994}. The code models the motion of one gravitationally dominant object (Sun) and $N$ massive objects (in our case: $N=8$) under the influence of their mutual gravity. Test particles move passively under the influence of the combined gravitational potential of the Sun and the planets. The algorithm handles close approaches by performing a time step regularisation by reducing the time step by a factor of 10 at 3.5 $R_{\rm Hill}$ and by another factor of 3 at 1 $R_{\rm Hill}$}

\rev{In order to check for impacts during the simulation the RMVS integrator checks if a test particle and planet are having, or will have within the next time step, an encounter such that the separation distance, $r$, is less than some critical radius, the boundary of the encounter region. If this is true the particle is discarded from the simulation. In order to catch all collisions Swift checks for intersections with planets and switches to an adaptive time step mode.} We are confident to set up the “baseline” time step in our models to 1 day in order to well resolve the 88-day orbit of Mercury and those of Mercury crossers. 

Test particles are removed from the simulation once they collide with a planet or with the Sun. Moreover, test particles are considered ejected from the Solar System and discarded when they exceed a user-provided heliocentric distance. These values are set separately for asteroid and comet simulations (see Subsections~\ref{method_asteroids} and \ref{method_comets}).

\subsection{Asteroids}\label{method_asteroids}

To model asteroid orbits we have used the MPCORB catalog from February 2017 (epoch K172G). The catalog contains orbital elements for a total of 730,272 asteroids. We culled inaccurate orbits based on single-epoch data, leaving us with 618,078 asteroids to model. We performed our simulations for 10~Myr forward in time in order to get a sufficient number of impactors while avoiding depletion of the impactor orbits. After reaching a heliocentric distance of 1,000~au, asteroids were considered ejected from the Solar System and discarded. 

\begin{figure}
\centering
\includegraphics[width=1.\linewidth]{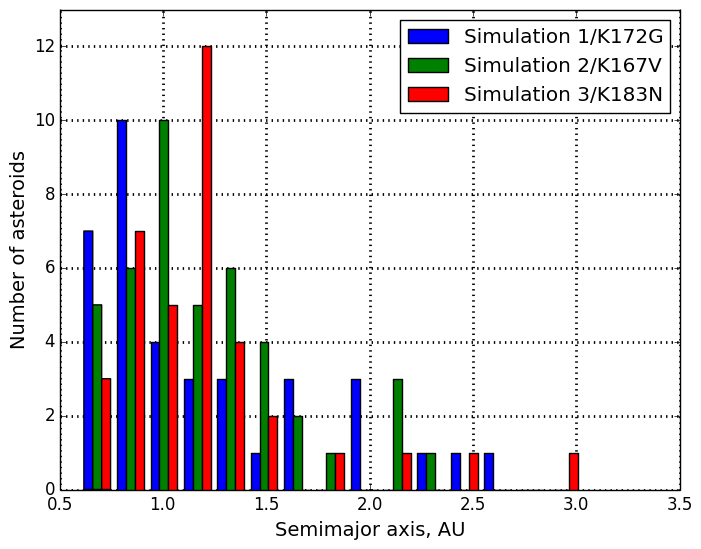}
\caption{Orbital distribution of the Mercury impacting asteroids. Blue bins correspond to the primary simulation, green bins to the first crosscheck simulation and red bins to the second crosscheck simulation. K172G, K167V and K183N stand for February 2017, July 2016, March 2018, which are the corresponding epochs for each simulation .}
\label{fig2:1}
\end{figure}

At the end of the simulation 12,480 asteroids were discarded from the system. 6,719 asteroids collided with the Sun, 4,609 asteroids were ejected and 1,152 asteroids collided with the planets. Among the latter, 37 asteroids collided with Mercury, corresponding to an average Mercury impact rate of 3.7 impacts per Myr. In Figure\ \ref{fig2:1}, \rev{in blue (Simulation 1/K172G)} we show the semimajor axes of all 37 impacting asteroids at the start of the simulations. They are distributed between 0.5~au and 3.5~au (Mercury's semimajor axis is 0.387~au) with a peak near 0.7~au.

As a crosscheck, we have performed two more asteroid simulations using different initial conditions. For the first crosscheck simulation we have used asteroid orbital elements from MPCORB as of July 2016 (epoch K167V, \revrev{618,078 asteroids}) together with planetary positions for the same date, while for the second crosscheck simulation we have used MPCORB as of March 2018 (epoch K183N, \revrev{645,245 asteroids}) for the asteroid orbital elements together with the corresponding planetary positions. \revrev{Both crosscheck simulation samples were cleaned using the same removal condition as for K172G.} 43 and 37 asteroids impacted Mercury during the first and second crosscheck simulation, respectively, consistent with our primary result of 37 impactors \rev{within the Poisson noise}. The individual asteroids that impact in the first simulation do not impact in the crosscheck simulations and vice versa, which means that our results are valid in a statistical sense\rev{,} only. We do not identify individual impacts, but we do characterize the population of Mercury impactors.

From here on we will refer to the primary simulation as Simulation~1, to the first crosscheck simulation as Simulation~2 and to the second crosscheck simulation as Simulation~3.

\begin{figure}
\centering
\includegraphics[width=1.\linewidth]{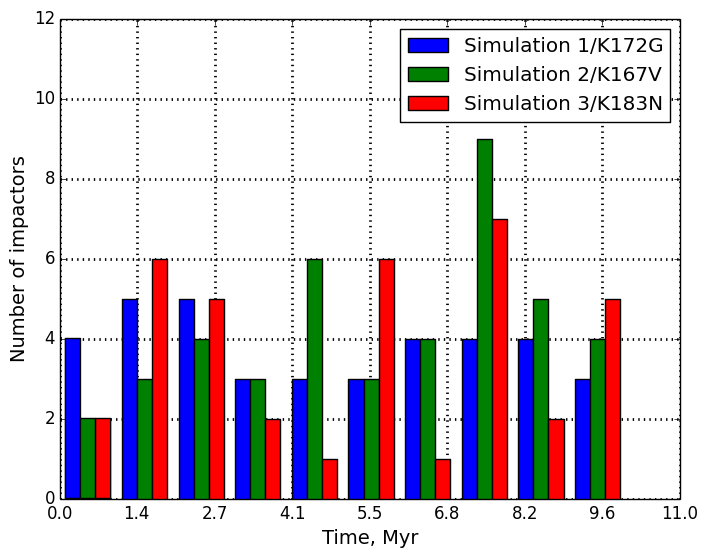}
\caption{Number of asteroids impacting Mercury per 1 Myr. Blue, green and red colors correspond to the primary and the two crosscheck simulations, respectively.}
\label{fig2:2}
\end{figure}

We analyzed the number of asteroids impacting Mercury after each Myr of the simulation as shown in Fig.\ \ref{fig2:2}. In this way we studied the time evolution of the Mercury impactors. We expect the number of impactors to be proportional to time. Any significant violation of our assumptions (e.g., depletion of asteroids on impactor orbits or Yarkovsky drift) would result in a deviation from this. We did find the \rev{total} number of impactors to increase steadily, within the $\sqrt{N}$ Poisson noise, justifying our assumptions.

\subsubsection{Asteroid collisional probabilities}\label{method_asteroids_probabilities}

\revrev{
Apart from recording direct impacts of asteroids in our simulations on Mercury we employ a more statistical approach to estimate the intrinsic collision probabilities of our asteroid populations with Mercury. Every 1 Myr we take the recorded orbital elements of the simulated asteroids $(a,e,I,\omega)$ and use two methods to calculate the collision probability of each asteroid with \revrev{the four terrestrial planets}: (1) \citet{Kessler1981} method, and (2) \citet{Pokorny2013} method. These methods calculate the volume of space swept by the target and the impactor assuming secular evolution of $(\Omega, \omega)$, and calculate the collision probability per unit time of two objects of interest. \citet{Kessler1981} requires the $(a,e,I)$ to be constants and $(\Omega,\omega)$ to precess uniformly, whereas \citet{Pokorny2013} includes Kozai-Lidov oscillations caused by a single perturbing body on a circular orbit, but requires the target body to be on a coplanar orbit. While neither of these approaches is perfectly reflecting the situation in our simulations, they provide valuable estimates for the asteroid/comet ensemble interaction with Mercury and other terrestrial planets.  }

\revrev{
For each planet both methods incorporate the effect of gravitational focusing, where the planet's effective cross-section is}
\begin{equation}
    \sigma = \pi (r_\mathrm{pl} + r_\mathrm{imp})^2 \left[ 1+ \frac{V_\mathrm{esc}^2}{V_\mathrm{rel}^2+\epsilon^2}\right],
\end{equation}
\revrev{
where $r_\mathrm{pl}$ is the radius of the target planet, $r_\mathrm{imp}$ is the radius of the impactor, $V_\mathrm{esc}$ is the escape velocity from the target planet, $V_\mathrm{rel}$ is the relative impact velocity between the planet and the target, and $\epsilon = 0.1$ km s$^{-1}$ is the softening parameter used to avoid singularities \revrev{for slow impactors}. We assume that impactors are much smaller than targets $r_\mathrm{pl}\gg r_\mathrm{imp}$, thus we keep $r_\mathrm{imp}=0$.}

\begin{figure}
\centering
\includegraphics[width=1.2\linewidth]{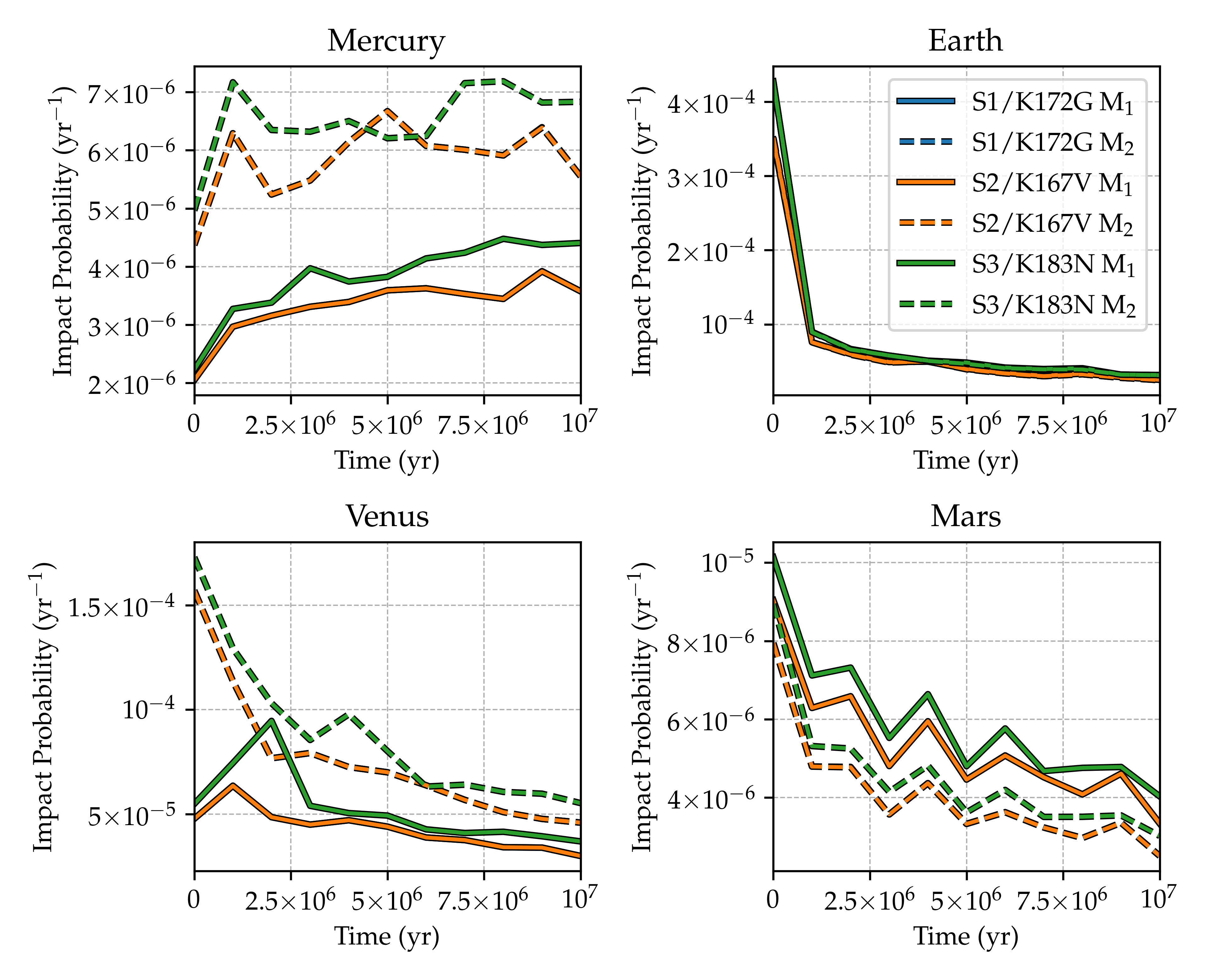}
\caption{\revrev{Impact probabilities of asteroids with the four terrestrial planets using (1) \citet{Kessler1981} method (solid lines; M1), and (2) \citet{Pokorny2013} method (dashed lines; M2). Our three simulation epoch color coded: K172G (blue), K167V (orange), K183N (green). Note that K172G and K167V provide almost identical results $(<0.1\% \mathrm{~difference})$.}}
\label{fig2.1.1:1}
\end{figure} 

\revrev{
For each simulated object is at every recorded time step we calculate the collision probability $\mathcal{P}$ using methods (1) and (2) and then sum $\mathcal{P}$ for all objects in our simulation which gives us the total collision probability with each planet per unit of time, see Figure~\ref{fig2.1.1:1}. Then we multiply the total collision probability with the recording time step (1 Myr) and get the total number of impacts in the recorded time period for methods 1 and 2 resulting in the harmonic mean asteroid impact probability of 4.41 impacts per Myr, which is 19\% more that the number of direct impacts.}

\revrevrev{
Note that the Mercury rate is constant/growing, as seen on Figure~\ref{fig2.1.1:1}, while the rate decreases with time for Venus, Earth and Mars. The difference in rates is caused by the fact that during the integration the simulated asteroids are injected into regions with the apocenter distance Q$<$0.90, where the asteroids are not currently surveyed very well. Only 10 asteroids with Q<0.9 are currently known. We do not consider mechanisms that could replenish the currently observed NEO population (i.e. asteroids interacting mainly with Venus, Earth, and Mars in our simulations), but we naturally have a mechanism (close encounters with terrestrial planets), that can inject asteroids into low Q orbits. In our simulations we start with 4 objects with Q$<$0.9 au and after 10 Myr, at the end of the simulations, we have 256 objects with Q$<$0.9 au.}

\subsection{Comets}\label{method_comets}

In order to model cometary orbits we have used the MPCORB catalog from November 2016 which contains 893 comets. Of these, we have used  879 comets whose orbits are calculated for the current epoch (removing comets for which the perturbed solutions were absent). As described in \citet{Frantseva2018}, the initial cometary set is too small to draw statistically valid conclusions, therefore we have replaced each comet with 5000 synthetic "clone" comets with randomized angular orbital elements (\rev{the semimajor axis $a$, eccentricity $e$, and inclination $i$ values were kept as of the actual comets}). Moreover, Mercury's radius was inflated by a factor of 50. The simulation was performed for 0.1 Myr forward in time; this simulation length was chosen to not significantly deplete the comet population over the course of our simulations. A comet was considered discarded from the simulation if it had collided with a planet, with the Sun, or when it reached a heliocentric distance of 125,000~au. \rev{This large} threshold value of the heliocentric distance is required by the high eccentricity of some comets.

At the end of the simulation, 76,261 synthetic comets were discarded from the simulation including 3,651 objects that had impacted Mercury.  Accounting for cloning of the comets, the inflation of Mercury's radius and the duration of our simulation, this corresponds to $3,651/5,000/2,500/0.1 = 0.0029$ impacts per Myr.

Like in the case of asteroids, we estimated the uncertainty in our comet result through a crosscheck simulation for a different epoch, November 2017. During the crosscheck simulation 3,684 comets impacted Mercury, consistent with our primary result of 3,651 impactors within the $\sqrt{N}$ Poisson noise.

\subsubsection{Steady-state cometary populations and their collisional probabilities}\label{method_coomets_probabilities}

\begin{figure}
\centering
\includegraphics[width=1.2\linewidth]{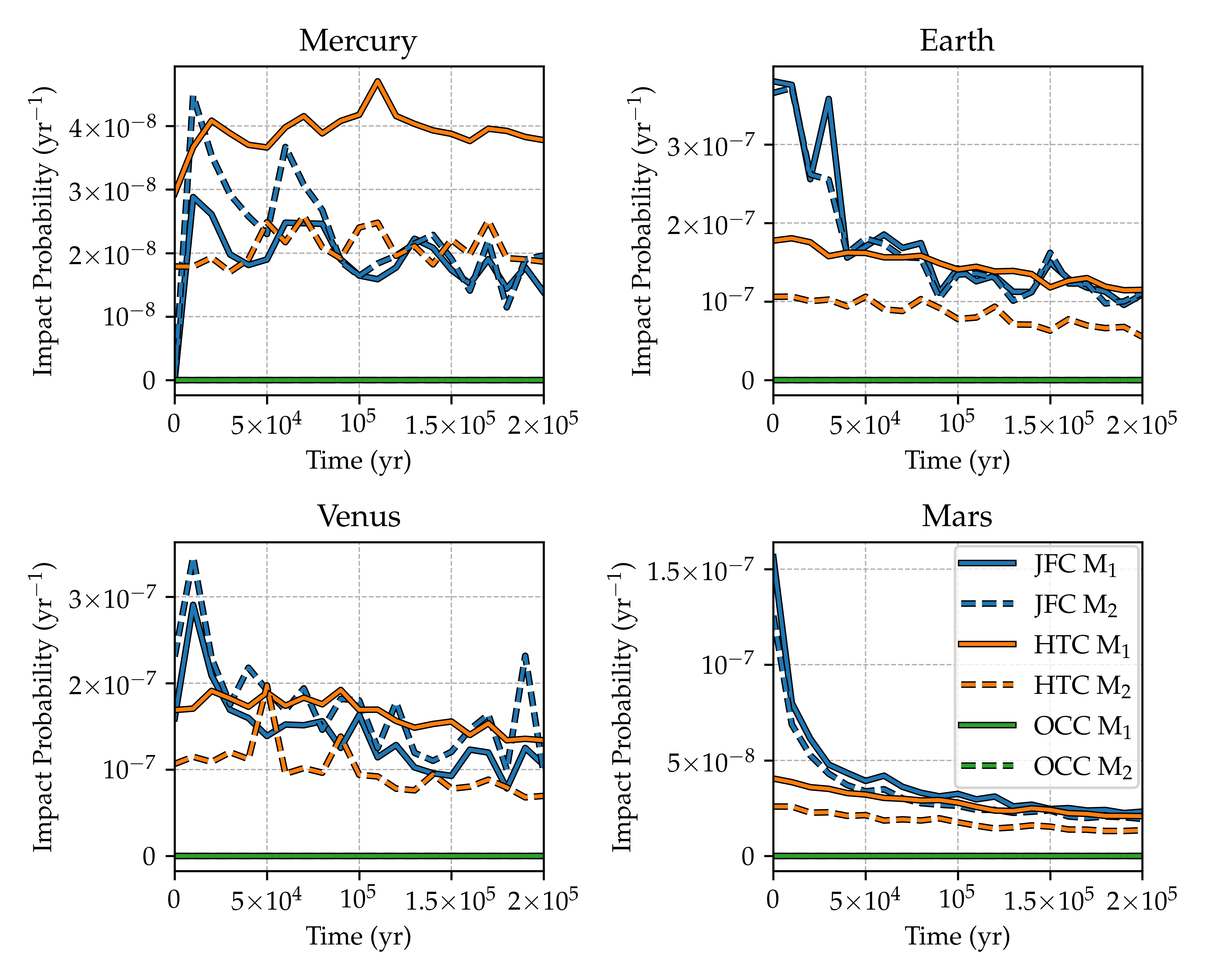}
\caption{\revrev{Impact probabilities of JFCs, HTCs and OCCs (10,000 synthetic orbits per population) with the four terrestrial planets  using (1) \citet{Kessler1981} method (solid lines; M1), and (2) \citet{Pokorny2013} method (dashed lines; M2). The three cometary populations are color coded: JFCs (blue), HTCs (orange), OCC (green). The OCC flux is 3-5 orders of magnitude smaller than that of JFCs and HTCs. }}
\label{fig2.2.1:1}
\end{figure}

\revrev{Besides comet simulations based on the MPCORB catalogue, we have also performed comet simulations based on the steady-state models for orbital element distributions for Jupiter Family comets (JFCs), Halley-type comets (HTCs), and Oort cloud Comets (OCCs) as described in \citet{Nesvorny2017} and \citet{Vokrouhlicky2019}. For the Jupiter Family comets we used the cumulative orbital distributions of ecliptic comets with orbital periods $P < 20$ yr, Tisserand parameter with respect to Jupiter $2 < T_{J} < 3$ and perihelion distance $q < 2.5$ au as in Figure~7 in \citet{Nesvorny2017}. For the Halley-type comets we used the cumulative orbital distributions with semimajor axes $10 < a < 20$ au, Tisserand parameter with respect to Jupiter $T_J < 2$ and perihelion distance $q < 2$ au as in Figure~11 in \citet{Nesvorny2017}. For the Oort cloud comets we have used the cumulative distributions of perihelia and cosine of ecliptic inclination with perihelion distance $q \leq 5$ au as in Figure~20 in \citet{Vokrouhlicky2019}. Based on the aforementioned distributions we have generated 10,000 comets for each of the populations and ran simulations similar to the ones performed with the asteroids extracted from MPCORB. The resulting impact probabilities with the four terrestrial planets are shown in Figure~\ref{fig2.2.1:1}. Because of their extremely eccentric orbits, OCCs have negligibly small impact probabilities compared to JFCs and HTCs. If we take the harmonic mean cometary impact probability of 10,000 synthetic comets (JFCs and HTCs) $\mathcal{P} = 2.34\times 10^{-8}$ yr$^{-1}$ then we get approximately 0.0021 cometary impacts per Myr assuming the same number of comets (879) as in Sec. \ref{method_comets}, which is 29\% smaller than the number we obtained by simulating the currently known cometary orbits.}

\section{Interplanetary Dust Particles}\label{method_idp}

We use the ZoDy model \citep{Nesvorny2010,Nesvorny2011a} to compute the accretion of IDPs on Mercury. In the ZoDy model, IDPs of different sizes are released from asteroids and comets. Their orbital evolution is followed by an efficient $N$-body integrator \citep[{\it Swift},][]{Levison1994} until they are ejected from the Solar System, or impact the Sun or a planet. The $N$-body integrations include the gravitational effects of the Sun and the planets.  Due to the small \rev{sizes} of the dust particles, non-gravitational forces such as radiation pressure, Poynting-Robertson and solar wind drag are also accounted for. 
The orbit of each particle is saved at 1,000 yr time intervals, thus allowing us to construct a steady state distribution of orbits for individual sizes. For example, for particles from Jupiter-family comets (JFCs), which probably are the most important source of IDPs in the inner Solar System (see below), we follow the orbital evolution of particles with 28 different sizes from 1~$\mu$m to 10~mm. Each set of integrations included 10,000 particles of the same size. 

In the absence of mutual collisions between particles, the size distribution of particles in space would be defined by their {\it initial} size distribution in the source and by their dynamical lifetimes. In the current ZoDy model, the initial size distribution is parametrized by a broken power law with a break at diameter $d^*$, and differential power law indices $\gamma_{\rm S}$ for small particles and $\gamma_{\rm L}$ for large particles. Motivated by the comparison of the ZoDy model with the Long Duration Exposure Facility (LDEF) satellite measurements \rev{\citep{Love1993,Nesvorny2011a}}, we use $d^*\simeq200$ $\mu$m, $\gamma_{\rm S}\simeq-2$ and $\gamma_{\rm L}\simeq-5$. In this case, both the mass and emission cross-section of IDPs are dominated by $d\simeq200$ $\mu$m particles, implying that the mass in micrometeoroids accreted by planets is also dominated by $d \simeq200$~$\mu$m particles. 

Once the initial size of the particles is fixed, the orbital integrations are used to compute the size distribution of particles at any location in the Solar System. To do this properly, we account for disruptive collisions between particles, which act to destroy parent particles and generate fragments. It is difficult to fully account for the collisional cascade in an $N$-body integrator because $N$ can increase enormously when new particles are added. We therefore adopt a standard approach to this problem \citep[e.g.,][]{Grun1985}, where the collisional lifetime, $\tau_{\rm c}$, is defined as a function of size and orbit, and used to decide whether a particle is removed. This approach ignores fragments generated by disruptive collisions. A Poisson random number generator is used to determine whether a particle is disrupted (depending on its size and orbital history). 

The orbital dependence of $\tau_{\rm c}$ can be determined from first principles \citep[see][for a discussion]{Nesvorny2011b}. The size dependence, however, is \textit{a priori} unknown and must be treated as a free parameter  (or a set of free parameters). In our past work, we used different prescriptions for $\tau_{\rm c}(d)$, including cases from Gr\"un et al. (1985), $\tau_{\rm c}(d)=constant$, and other dependencies where $\tau_{\rm c}$ increases with $d$. To fit the meteor data \rev{on Earth}, for example, we found that $\tau_{\rm c} \gtrsim 10^5$ yr for $d>100$ $\mu$m (for a reference orbit with $a=1$ au and $e=i=0$). 

\citet{Nesvorny2010} used the ZoDy model to determine the relative contribution of asteroid and cometary material to the Zodiacal cloud and the mass in IDPs accreted by the Earth. They found that the mid-infrared (MIR) emission from particles produced in the asteroid belt is mostly confined to within latitudes $b\lesssim30^\circ$ of the ecliptic. 
Conversely, the Zodiacal cloud has a broad latitudinal distribution such that strong thermal emission is observed even in the direction of the ecliptic poles. This shows that asteroidal particles can represent only a small fraction \citep[under 10\%, e.g.,][]{Carrillo-Sanchez2016} of the Zodiacal cloud emission. Their contribution to the mass accreted by planets can be larger than that \rev{10\%}, because asteroid IDPs move on low eccentricity and low inclination orbits, their velocities with respect to planets are lower, and their impact cross-section is therefore strongly enhanced by gravitational focusing. 

Based on a detailed comparison of the dynamical model with MIR observations by the Infrared Astronomical Satellite (IRAS) and the Cosmic Background Explorer (COBE), \citeauthor{Nesvorny2010}\ found that $\gtrsim$90\% of the zodiacal emission at MIR wavelengths comes from dust grains released by JFCs, and only $\lesssim$10\% comes from the long periodic comets. The total mass accreted by the Earth in JFC particles between diameters $D=5$ $\mu$m and 1 cm was found to be \rev{$\sim\,15\times10^{6}\,$kg yr$^{-1}$} (Nesvorn\'y et al. 2011a; factor of 2 uncertainty), which is a large share of the accretion flux measured by LDEF \citep{Love1993}. Based on these results we consider only IDPs from asteroids and JFCs for our calculations. \revrev{All other populations, such as long period comets \citep{Wiegert2009,Pokorny2014} and the Kuiper belt, are ignored due to their high impact velocities with the Hermean surface \citep[$>80$ km/s, ][]{Pokorny2017,Pokorny2018} - at these velocities any water content is likely lost to impact vaporization/ionization \citep[e.g., ][]{Cintala1992}.} In Section \ref{water_idp}, we use the ZoDy model for asteroid and JFC particles to determine the impact flux of IDPs on Mercury \citep[also see][]{Pokorny2018}. 

\section{Water delivery rates}\label{water_delivery}

In \rev{Sections \ref{method_asteroids} and \ref{method_comets}}, we estimated that Mercury suffers 0.0029 comet impacts per Myr and 3.7 asteroid impacts per Myr. \revrev{Additionally, our statistical approach yielded 0.0021 cometary impacts per Myr and 4.41 asteroidal impacts per Myr. In this section we will use the direct impact rates and convert them into water delivery rates. Based on the described methods for the IDP flux we will estimate the corresponding water flux on Mercury.}  

\subsection{Asteroids}\label{water_asteroids}

Asteroids are known to be parent bodies of meteorites. We here focus on the carbonaceous chondrite meteorites, which are known to be water and organic rich. Their water content is $\sim10\%$ by mass, their carbon content is $\sim2\%$ by mass and their parent bodies are the \rev{C type} asteroids \citep{EOE2007,Sephton2002,Sephton2014}. By comparison, other taxonomic types contain negligible amounts of water; they are neglected. 

To calculate the water delivery rate of the asteroid impacts we use the statistical approach discussed in \citet{Frantseva2018}, where each impactor is assigned a probability $p_\text{C}$ of \rev{belonging to the C type} depending on its semimajor axis at the start of the simulations. The amount of water delivered by an asteroid depends on $p_\text{C}$, the asteroid mass $m_\text{Asteroid}$ and $f_\text{Water}$, the water content of carbonaceous chondrite meteorites:
\begin{equation}
\label{eq3:1}
M_{\rm Water} = p_\text{C} m_\text{Asteroid} f_\text{Water}.
\end{equation}
The water delivery rate equals the sum over all impactors, divided by the duration of our simulation.

The diameter $D$ of an asteroid is based on the $H$ magnitude \citep{Bowell1989} and geometric albedo $p_V$ \citep{Pravec2007}:

\begin{equation}
\label{eq3:2}
D = \frac{1329 \rm km}{\sqrt{p_V}} 10^{-H/5} 
\end{equation}
where $p_V = 0.06 \pm 0.01$ \citep[a representative value for \rev{C type} asteroids, see][Table 1]{DeMeo2013}. Mass follows from diameter adopting an average \rev{C type} mass density $\rho$ of $1.33 \pm 0.58$~g/cm$^3$ \citep[Table 3;][]{Carry2012}:
\begin{equation}
\label{eq3:3}
M = \frac{\pi}{6}  D^3 \rho.
\end{equation}

The probability of an asteroid being part of the \rev{C type}, $p_\text{C}$, is estimated based on the initial semimajor axis $a$ of an asteroid at the start of the simulation.
For impactors with $a<2$ au, we identify $p_\text{C}$ with  the "dark fraction" in the albedo distribution of NEOs as measured by WISE \citep[dark NEOs are predominantly asteroids of the \rev{C class}]{Wright2016}.  For the remaining impactors, we use the measured fraction of C types relative to all asteroids derived by \citet{DeMeo2013}, based on NIR spectroscopic surveys. \citeauthor{DeMeo2013} provide two sets of results, before and after debiasing against albedo-dependent survey efficiency, we use the latter. WISE is not subject to such a bias, therefore the C-type fraction for $a<2$ au need not be debiased.
We adopt $p_\text{C}$ values of 0.253 for $a<2$ au \citep{Wright2016}; 0.0612 for $2$ au $< a < 2.18$ au,  the \citeauthor{DeMeo2013} value for $a=2.18$ au. For $a>2.18$ au, we use C-type fractions specified for each 0.02 au bin in Figure 9 in \citet{DeMeo2013} \citep[see also Fig.\ 2 in][]{Frantseva2018}.
In practice, the majority of Mercury impactors has $a<2$ au, see Fig.\ \ref{fig2:1}.

\begin{table}
\begin{center}
\begin{tabular}{ l l m{2cm} l }
\hline
\rev{Simulation} & Epoch & \rev{Number of impactors} & Water flux, \rev{$10^{6}$ kg/yr}\\ 
\hline
\rev{1} & K172G, February 2017 & 37 & \rev{$0.021 \pm 0.009$} \\ 
\rev{2} & K167V, July 2016 & 43 & \rev{$0.023 \pm 0.009$} \\  
\rev{3} & K183N, March 2018 & 37 & \rev{$0.430 \pm 0.190$} \\
\hline
\end{tabular}
\end{center}
\caption{For each of our three asteroid simulations we report the epochs at which they were initialized, the resulting numbers of Mercury impacting asteroids, and the corresponding water fluxes.}
\label{table:1}
\end{table}

Using Equation~\ref{eq3:1}, we estimate a rate of water delivery to Mercury due to asteroid impacts of $0.021 \pm 0.009 \times 10^6$ kg/yr, averaged over 10 Myr. The uncertainty is calculated combining the known uncertainties of $H$ magnitude, albedo, and mass density (the uncertainty in $p_\text{C}$ and water content are neglected). 

Using the same method, we also analyzed the results of the two crosscheck simulations, Simulations 2 and 3, resulting in a mean value for the water delivery rate of $0.023 \pm 0.009 \times 10^6$ kg/yr and \rev{$0.430 \pm 0.190 \times 10^6$ kg/yr,} respectively.  While Simulation 2 is in good agreement with Simulation 1, Simulation 3 differs by a factor of nearly \rev{20}. We attribute this discrepancy to the different size-distribution of the Mercury impactors: the third simulation contained several relatively big impactors of 2-5~km in diameter, while impactors in the primary and first crosscheck simulation were smaller. The largest few impactors dominate the mass budget, leading to large Poisson noise. \citet{Moses1999} report similar mass-flux differences between simulations due to the mass flux being dominated by the largest few impactors. The results of the three asteroid simulations are listed in Table \ref{table:1}. \rev{We adopt the harmonic mean of the three runs,}  \rev{$0.032 \times 10^6$ kg/yr} within a factor of four to five, as our final asteroid delivery rate. 

Our calculations are based on the present-day catalogs of asteroids and comets, which are known to be incomplete. They are missing the smallest objects that are too faint to be detected. Nevertheless, the largest asteroids will dominate the total flux while the contribution of the undiscovered asteroids is only $\sim$4\% as described in \citet{Frantseva2018}. Observational incompleteness of asteroid catalogs is therefore uncritical for our purposes. 

\subsection{Comets}\label{water_comets}

\rev{The two simulations in Section \ref{method_comets} agree on} the cometary impactor flux on Mercury is \rev{$\approx0.0029$} comets per Myr. In order to calculate a water delivery rate to Mercury we need to know the typical comet mass and water content. We assume the typical comet mass to be $3 \times 10^{13}$ kg following \citet{Swamy2010}. Various comet observations suggests the water content to lay in a range \revrev{between} 3\% and 90\% \citep{Gicquel2012,Huebner2002,Jewitt2004,Taylor2017} and 50\% from the comet nucleus modeling \citep{Prialnik2002}. For our calculations we adopted an average value of 50\% due to the diversity of the observed and modeled cometary water content. We estimate the water delivery rate to Mercury to be $0.044 \times 10^{6}$ kg/yr. 

As in the case of asteroids, our comet catalog is observationally incomplete, with a bias against small, faint objects.  We neglect that bias for the same reason as for asteroids.
Additionally, we are strongly biased against discovering comets with large orbital periods (due to the finite temporal baseline available).
Systematic surveys (aimed at discovering potentially hazardous asteroids, but discovering comets as bycatch) started in the 1990s, so for orbital periods up to 20+ years, our catalogs should be reasonably complete.  Those provide the bulk of the water retained on Mercury; longer-period comets hit at very large relative velocity, causing most water to be lost to space (see discussion below).
Therefore, observational incompleteness is as uncritical for comets as it is for asteroids.

\subsection{Interplanetary Dust Particles}\label{water_idp}

Using the methods described in Section \ref{method_idp}, we determine the ratio of IDP impact fluxes between the Earth (here used as a calibration point) and Mercury. For JFC particles, we find that the IDP flux on Earth should be $\sim$4 times larger than on Mercury, mainly owing to its larger impact cross-section (the Earth's cross-section is $\simeq$7 times larger than that of Mercury). For asteroid particles, the Earth should receive $\sim$15 times the IDP accretion of Mercury, a significantly higher ratio than the one obtained for  JFC particles. This difference is probably related to stronger gravitational focusing of asteroid IDPs by the Earth (asteroid IDPs have lower inclinations and lower eccentricities, thus lower impact velocities).

We now need to convert these ratios to absolute impact fluxes. For that we use a calibration from \citet{Nesvorny2011a}, who found that Earth accretes $\sim 2 \times 10^7$ kg/yr in JFC IDPs, which is about half of the total input measured by LDEF. If so, Mercury should receive $5 \times 10^6$ kg/yr of JFC IDPs. 

The computation of asteroid IDP flux on Mercury is more uncertain, because we do not have a reliable calibration of asteroid IDP flux on  Earth. For example, if we assume that the asteroid IDPs are responsible for roughly half of the IDP flux measured flux by LDEF, then we can estimate that Mercury should receive only $\sim 10^6$ kg/yr in asteroid IDPs (or $\sim 2\times 10^6$ kg/yr if asteroid IDPs are responsible for the full flux measured by LDEF). We will adopt $\sim 10^6$ kg/yr as the asteroid IDP flux to Mercury.  

We estimate the joint asteroid and comet IDP flux on Mercury to be $6 \times 10^6$ kg/yr. \rev{Given the uncertainties, our estimate is in good agreement with $4.4 \times 10^6$ kg/yr by \citet{Pokorny2018}, who used a similar model.} 

Some fraction of IDPs are anhydrous and some are hydrous. The values of these fractions have been studied in the past. It has been estimated that the hydrous IDPs can be from 1\% and up to 75\% of the total IDPs \citep{Engrand1999,Noguchi2002,Dobrica2010,Zolensky1992}. Here we will use $\sim$ 40\% as a conservative proportion of dust particles initially containing water. The water content of the hydrous dust particles was found to be from 1wt\% and up to 40wt\% \citep{Engrand1996}. We will use 20\% water content of the hydrous dust particles. 

All told, we estimate that dust particles deliver $\sim 0.5 \times 10^6$ kg/yr within a factor of $\sim 2$ of water to Mercury.

\section{Water survivability}\label{survivability}

In the previous section we converted \revrev{the} impact fluxes of asteroids, comets\rev{,} and IDPs to the corresponding water fluxes. We have shown that Mercury will receive \rev{$0.032 \times 10^6$ kg/yr} of water from asteroids, $0.044 \times 10^{6}$ kg/yr of water from comets, and \rev{$0.500 \times 10^6$ kg/yr} of water from IDPs. 

However, only a fraction of this water will be deposited at the poles. 
A (possibly large) fraction is lost immediately after impact, as impactor material evaporates, forming a supersonic plume, with some of its water content accelerated beyond escape velocity.  
The water that is retained by Mercury's gravity will migrate across the surface until it is caught at the poles; in the mean-time, some fraction of it is lost to dissociation. 

\subsection{Water loss at impact}

Asteroids and comets \rev{just like dust particles} hit the surface of Mercury at high relative velocity. \rev{All impactors including dust particles reach the surface essentially undecelerated by Mercury's tenuous exosphere, not even small dust grains feel Mercury's exosphere appreciably \citep{Ceplecha1998,Ceplecha2005}. } Impacts on planetary surfaces are a topic of extensive research \citep[see, e.g.,][for an overview]{Jutzi2015}. Impacts cause a (potentially large) fraction of the asteroid\rev{, comet and dust} material to reach velocities beyond escape velocity immediately after impact, so that material is lost. The fraction of retained impactor material is governed by the relative velocity between impactor and Mercury at the time of impact, the impact speed.  Fast impactors will lose most material to space, while material from slower impacts is retained more efficiently.  Impact velocity, in turn, depends chiefly on orbital parameters: high $e$ and high $i$ favor high impact velocity.

\begin{figure}
\centering
\includegraphics[width=.99\linewidth]{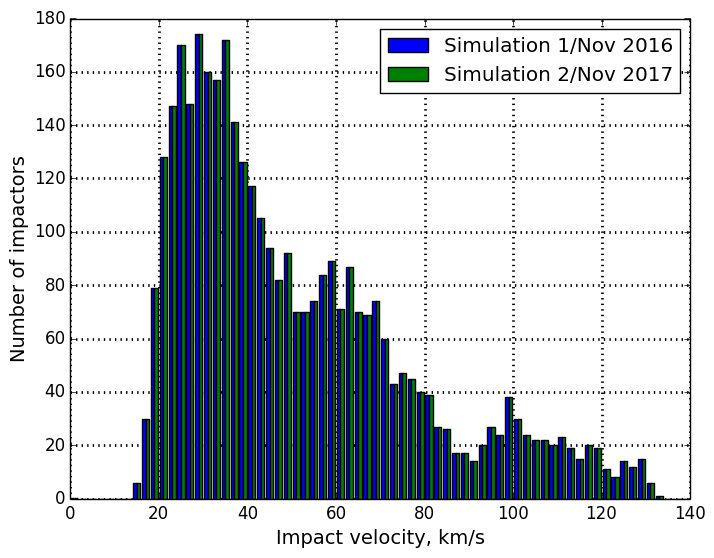}
\caption{Velocities at which comets impact  Mercury. Blue bins correspond to the simulation performed using data from MPCORB as of November 2016 and green bins correspond to the simulation performed using data from MPCORB as of November 2017; the distribution in impact velocity is indistinguishable.}
\label{fig5:1}
\end{figure}

\begin{figure}
\centering
\includegraphics[width=.99\linewidth]{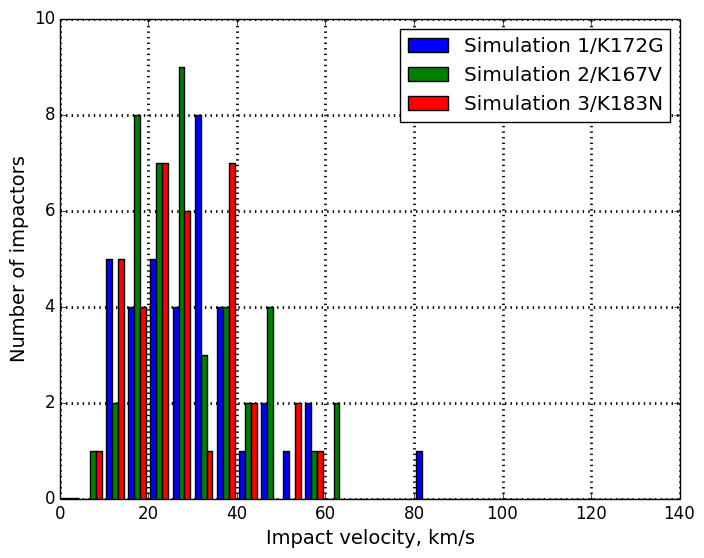}
\caption{Velocities at which asteroids impact Mercury. Blue bins correspond to the primary simulation, green bins correspond to the first crosscheck simulation and red bins correspond to the second crosscheck simulation.}
\label{fig5:2}
\end{figure}

\rev{For asteroids and comets} we estimate the relative velocity between impactor and Mercury from the RMVS output just before a test particle (asteroid or comet) is discarded, at the time step where the integrator recognizes that the test particle will impact.  That time step is much less than \rev{one} day before the time of impact, \revrev{and} the relative velocity obtained from it should be a good approximation for the actual impact velocity.
See Fig.\ \ref{fig5:1} for a histogram of \revrev{the} impact velocities of comets, \revrev{and} Fig.\ \ref{fig5:2} for asteroids. Most of the particles are seen to impact at velocities of 20-40 km/s.

\begin{figure}
\centering
\includegraphics[width=.99\linewidth]{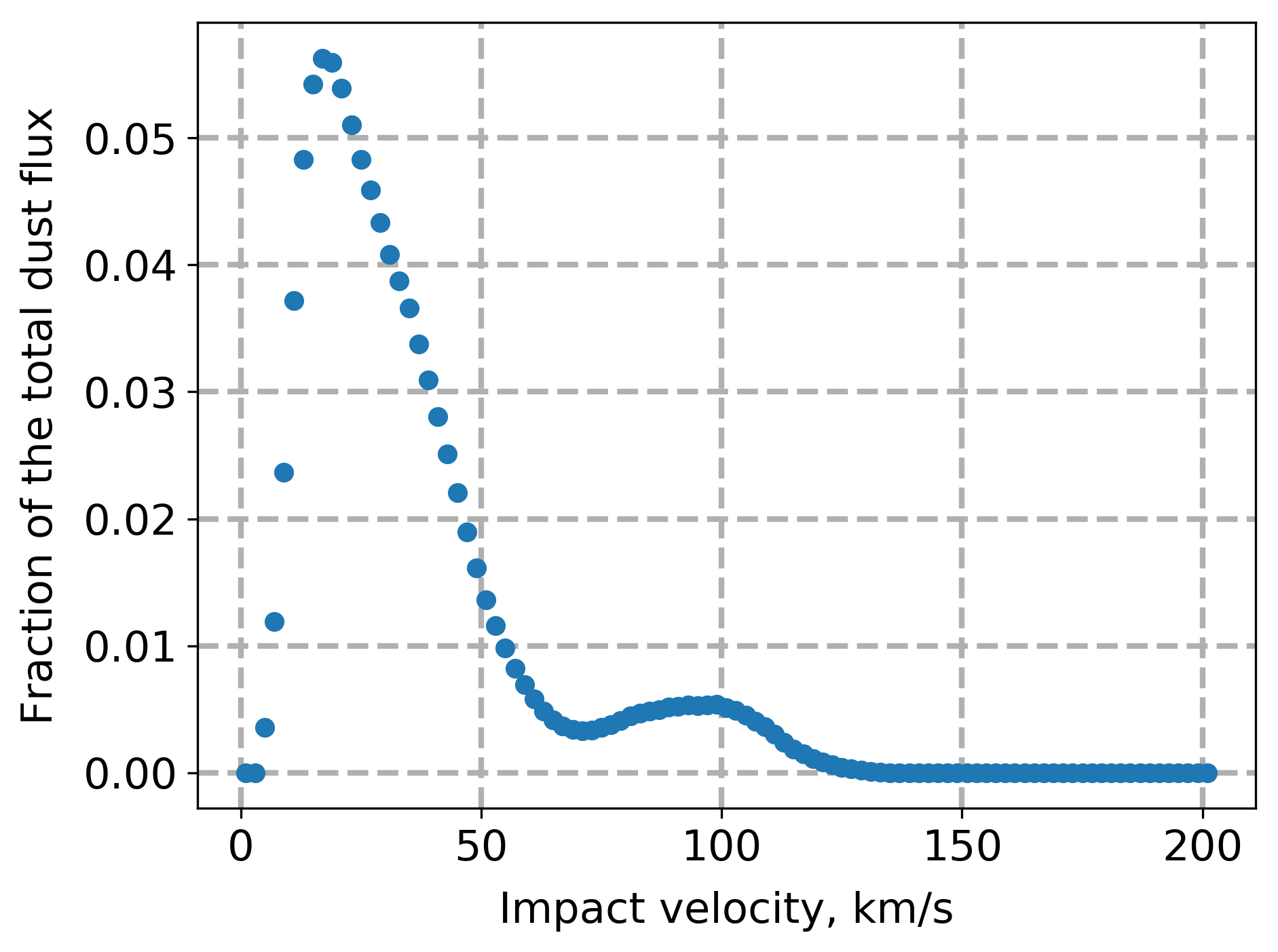}
\caption{\rev{Impact velocity distribution of the total dust flux on Mercury \citep[derived from][Figure 22]{Pokorny2018}.}}
\label{fig5:3}
\end{figure}

\rev{For the interplanetary dust particles we use a distribution of dust mass flux as function of impact velocity from \citet[their Figure 22]{Pokorny2018}. In particular, we take their mass influx distribution as a function of Mercury's true anomaly and the impact velocity. We averaged the distribution  over Mercury's orbital period as function of impact velocity and normalise it by the total daily dust flux. Fig.~\ref{fig5:3} shows the resulting distribution of the total dust flux on Mercury as function of impact velocity.}

\begin{figure}
\centering
\includegraphics[width=.99\linewidth]{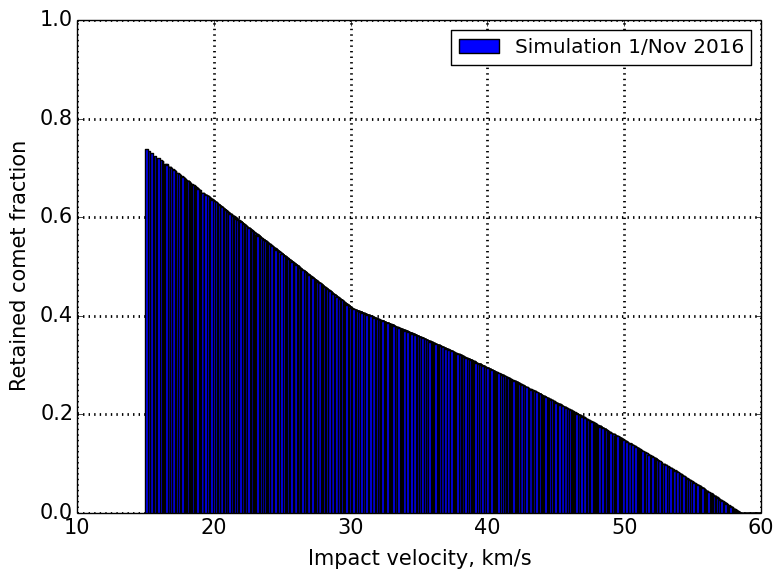}
\caption{Retained mass fraction of each comet impacting Mercury in the first comet simulation. For impact velocities larger than 60 km/s, all impactor material (and hence all water) is lost to space. The retained mass fraction of the impacting comets in the second simulation is indistinguishable from the first simulation and is not plotted here.}
\label{fig5:4}
\end{figure}

\begin{figure}
\centering
\includegraphics[width=.99\linewidth]{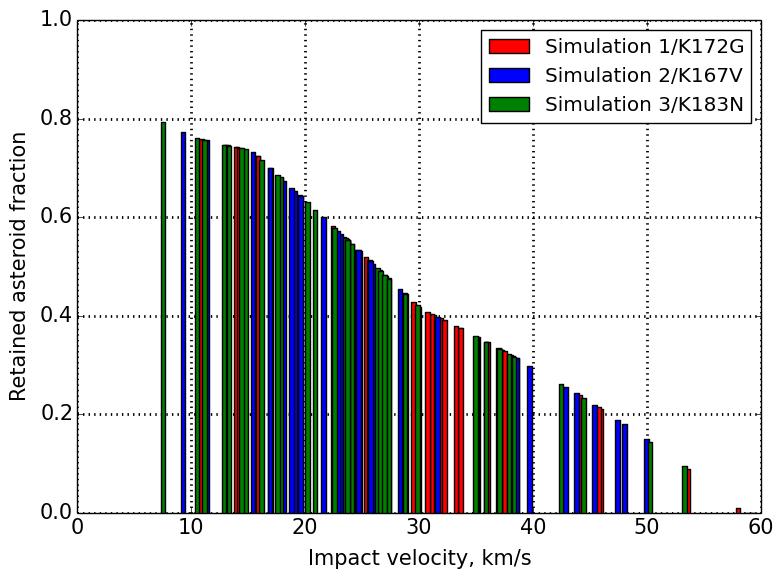}
\caption{Retained mass fraction of each asteroid impacting Mercury in all three simulations. For impact velocities larger than 60 km/s, all impactor material (and hence all water) is lost to space.}
\label{fig5:5}
\end{figure}

\begin{figure}
\centering
\includegraphics[width=.99\linewidth]{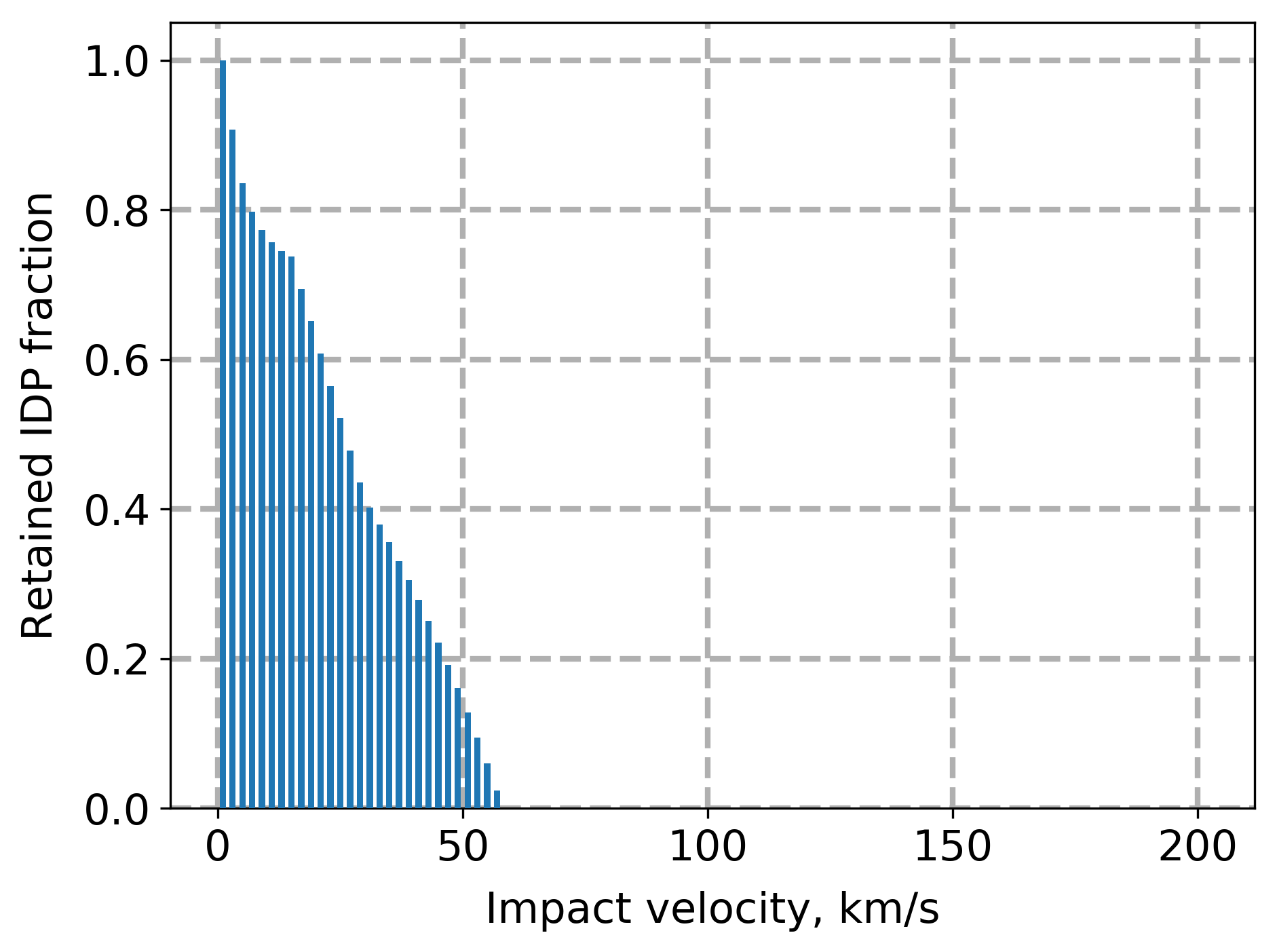}
\caption{\rev{Retained mass fraction of interplanetary dust particles impacting Mercury. For impact velocities larger than 60 km/s, all impactor material (and hence all water) is lost to space.}}
\label{fig5:6}
\end{figure}

To estimate the retained mass fraction \rev{of asteroids, comets and dust particles} upon impact, scaling laws have been established based on comparison between high-speed impact experiments and numerical simulations.  We base our analysis on the formalism developed by \citet{Svetsov2011}, which is accurate enough for our purposes. To estimate the fraction of the escaped projectile mass, $\frac{m_p}{m}$, for the impactors with low impact velocities $V \leq 15$ km/s, we use their Equation 8:

\begin{equation}
\label{eq5:1}
\frac{m_p}{m} = 1 - (0.14 + 0.003V) \ln v_{\rm esc} - 0.9V^{-0.24}, 
\end{equation}

while for high impact velocities $V \geq 30$ km/s we use their Equation 9:

\begin{equation}
\label{eq5:2}
\frac{m_p}{m} = \exp\left(\left(0.0015V - 0.2\right)v_{esc} + 0.0125V - 0.25\right),  
\end{equation}

where all velocities are in units of km/s and $v_{esc} = 4.25$ km/s is Mercury's escape velocity. In the velocity range between 15 and 30 km/s, we use a linear interpolation between Equations \ref{eq5:1} and \ref{eq5:2}. \rev{Figures \ref{fig5:4}, \ref{fig5:5} and \ref{fig5:6} illustrate the retained mass fraction, $1 - \frac{m_p}{m}$,  for each comet and asteroid that impacts  Mercury in our simulations and for the total dust flux.} For impact velocities up to 45 km/s, at least 20\% of material is retained on the surface. For impact velocities beyond $60$ km/s, all impactor material is lost to space.

Taking into account the retained fraction of each comet, we estimate the total water flux at Mercury to be $0.0105 \times 10^6$ kg/yr for the primary comet simulation and $0.0106 \times 10^6$ kg/yr for the crosscheck comet simulation. Our final adopted cometary retained water flux is $0.0105 \times 10^6$ kg/yr. 
For the asteroids, we estimate the water flux at Mercury to be $0.008 \times 10^6$ kg/yr, $0.013 \times 10^6$ kg/yr and $0.074 \times 10^6$ kg/yr for the asteroid Simulation 1, Simulation 2\rev{,} and Simulation 3 correspondingly as shown in Table \ref{table:2}. Our final adopted asteroid retained water flux is \rev{the harmonic mean} of the range spanned by our results of $0.014 \times 10^6$ kg/yr within a factor of three.  
\rev{For the interplanetary dust particles, our estimated value of the retained water flux is $0.207 \times 10^6$ kg/yr.}

\subsection{Water loss during migration to cold \rev{traps}}

Impacts of asteroids, comets and IDPs are spread homogeneously over \rev{all latitudes}
of the planet.%
\footnote{\rev{Based on MESSENGER observations, \citet{Fassett2012} demonstrated that impact craters on Mercury show a non-uniform distribution in longitude.  While we do not attempt to explain that finding, it does not influence our analysis in this section.  A hypothetical inhomogeneity in latitude, which is not found, would.}} Part of the water delivered by the impacts will reach the polar regions through migration processes. \citet{Butler1993,Butler1997} have performed a Monte Carlo molecule migration simulation to study which fraction of volatiles will be captured in the cold traps of Mercury. Their simulations show that $\approx 5-15$\% of the water molecules will survive to reach the cold polar regions before being dissociated by solar ultraviolet radiation. Combining the migration rates with our retained water fluxes it follows that \rev{$1 \times 10^3$ kg/yr} of \rev{asteroid-borne} water will be trapped in the permanently shadowed regions, analogously \rev{$1 \times 10^3$ kg/yr} of comet-borne water and \rev{$16 \times 10^3$ kg/yr} of IDP-borne water. Our final adopted values are the \rev{harmonic means of the ranges} spanned by our results \revrev{within} a factor of several. 

\begin{table}
\begin{center}
\begin{tabular}{ l l l l}
\hline
 & Incoming flux & Retained flux & Post-migration flux  \\ 
\hline
Simulation 1 & $0.021 \pm 0.009 \times 10^6$ & $0.008 \times 10^6$ & $(0.4 - 1.2) \times 10^3$ \\ 
Simulation 2 & $0.023 \pm 0.009 \times 10^6$ & $0.013 \times 10^6$ & $(0.7 - 2.0) \times 10^3$ \\ 
Simulation 3 & $0.430 \pm 0.190 \times 10^6$ & $0.074 \times 10^6$ & $(3.7 - 11.1) \times 10^3$ \\
\hline
Final value  & \rev{$0.032 \times 10^6$}     & \rev{$0.014 \times 10^6$} & \rev{$1.05 \times 10^3$} \\
\hline
\end{tabular}
\end{center}
\caption{For each of our three asteroid simulations, we list the corresponding water fluxes before impact (column 1), after post-impact ejection (column 2), and the water flux that successfully migrates to the polar cold traps (column 3).  All values are in units of kg/yr.  In the last row, we provide our final adopted values of each column, where we adopt \rev{the harmonic mean of the ranges} spanned by our results; uncertainties are a factor of $4-5$ in column 1, a factor of $\sim$ 3 in column 2, and a factor of $\sim$ 5 in column 3.}
\label{table:2}
\end{table}

\section{Discussion}\label{discussion}

Asteroids, comets\rev{,} and IDPs are possible exogenous sources of water on Mercury. In the previous sections, we find that asteroids, comets and IDPs deliver \rev{$\sim 1 \times 10^3$ kg/yr, $\sim 1 \times 10^3$ kg/yr and $\sim 16 \times 10^3$ kg/yr within a factor of several} of water to Mercury's polar regions, respectively. IDPs are the dominant source of water, \revrev{while asteroids and comet both deliver about an order of magnitude less.} 

IDP fluxes on Mercury have also been calculated by \citet{Moses1999}, \citet{Borin2017} \rev{and \citet{Pokorny2018}}. Our estimate for the total IDP influx on Mercury, $6 \times 10^6$ kg/yr, is a factor of 2 lower than the value of \citet{Moses1999} of $10^7$ kg/yr, a factor of 40 lower than the value of \citet{Borin2017} of $2 \times 10^8$ kg/yr \rev{and in good agreement with the estimate of $4.4 \times 10^6$ kg/yr by \citet{Pokorny2018}}. The difference relative to \citet{Moses1999} can easily be explained by the different adopted values for the measured IDP rate on Earth, reconciling our results. \revrev{The difference relative to \citet{Borin2017} can be attributed to the fact that our results were calibrated using \citet{Love1993}, while \citet{Borin2017} calibrated their results using \citet{Cremonese2012}. Also, \citet{Borin2017} do not include mutual meteoroid collisions that are important for Hermean impactors. Thus their model should be treated as the upper limit. As shown in \citet{Pokorny2018} collisions play a major role for all IDPs with diameters $D>100~\mu$m effectively removing a significant portion of larger asteroidal and JFC meteoroids from the pool of potential Hermean impactors.}

Our delivery rates due to asteroid and comet impacts can be compared to \revrev{the} results by  \citet{Moses1999}. They found that asteroids would deliver \rev{$(0.4 - 20) \times 10^{13}$ kg} of ice in 3.5 Gyr, \rev{consistent with} our estimate of \rev{$\sim 0.4 \times 10^{13}$ kg}. Our results should be more accurate than those of \citet{Moses1999}, chiefly because they are based on the known distribution of asteroid orbits as of 2018 rather than an extrapolation of the 1994 dataset. Furthermore, our probabilistic approach to define water-rich asteroids is more advanced than the flat 5\% water content assumed by \citet{Moses1999}. 

For comets, \citet{Moses1999} treat Jupiter-family comets and Halley-type comets separately.  For the former, they estimate a delivery rate of \rev{$(0.1-200) \times 10^{13}$ kg} in 3.5 Gyr, and \rev{$(0.2-20) \times 10^{13}$ kg} in 3.5 Gyr for the latter. Our estimate for water delivered by the entire comet population is \rev{$\sim 0.4 \times 10^{13}$ kg} in 3.5 Gyr.  While these results do agree at the lower range of the quoted uncertainty intervals, much larger delivery rates are allowed by \citet{Moses1999} than by our results. We attribute this chiefly to their large model-dependent correction factors accounting for the observational incompleteness of comets.  Compared to \rev{them}, we benefit \revrev{from} 20 years of systematic sky surveys looking for asteroids (and occasionally discovering comets as bycatch). This should largely eliminate observational incompleteness for large Jupiter-Family Comets with their orbital periods of 20 years or less; correspondingly, we should estimate the delivery rate due to Jupiter-Family Comets rather accurately (note that delivery is dominated by the largest impactors). Delivery due to longer-period comets would be underestimated in our model, \rev{on the other hand}. \rev{However}, long-period comets impact at large relative \rev{velocities}, causing most water to be lost to space. \revrev{Our comet simulations based on the steady-state models for orbital element distributions for JFCs, HTCs, and OCCs \citep{Nesvorny2017,Vokrouhlicky2019} yielded similar result to the number we obtained by simulating the currently known cometary orbits.}

Based on the radar observations the total mass of water ice on Mercury's poles was calculated to be $4 \times 10^{13} - 8 \times 10^{14}$ kg \citep{Moses1999}. Using the MESSENGER observations it has been shown that the total mass of water on Mercury's poles is $2.1 \times 10^{13}$ to $1.4 \times 10^{15}$ kg \citep{Lawrence2013}, consistent with the radar result. MESSENGER found the surface area of the permanently shadowed regions around the north pole to be \rev{$(1.25 - 1.46) \times 10^{10}$ m$^2$} and \rev{$(4.3 \pm 1.4) \times 10^{10}$ m$^{2}$} around the south pole. Combining the neutron and radar data from the MESSENGER spacecraft, the water ice origin of the radar-reflective deposits was confirmed for the North polar region (NS was not sensitive to the South due to the spacecraft's eccentric orbit). The total amount of water in the deposits is calculated based on the assumption that the South pole deposits have the same composition \citep{Lawrence2013}. Additionally, estimates of the deposit mass, based on MESSENGER observations, assume that the deposits are between 0.5\,m and 20\,m deep. That layer depth follows from models of surface modification processes \citep{crider2005} and from models of the radar scattering \citep{Butler1993}, not from MESSENGER data directly.

We find that impacts of IDPs, asteroids, and comets deliver \rev{$\sim 18 \times 10^3$ kg/yr} of water to Mercury's poles. This is easily enough to explain the observational lower limit on the ice-layer thickness of $2.1 \times 10^{13}$ kg; delivery would take \rev{$\sim 1$ Gyr}. \rev{While our analysis does not rule out any other sources of water ice on Mercury, we do show that none are needed to explain the (lower limits on the) data available today. 
Over 3.5 Gyr impacts of dust, asteroids and comets would deliver up to $6 \times 10^{13}$ kg. More is allowed but not required by the data. Should evidence for more water ice than $6 \times 10^{13}$ kg be found in the future, this would necessitate additional water sources beyond impacts.}

\revrev{When converting the number of asteroid impacts to the water flux we focused on the C type asteroids parent bodies of chondrite meteorites, assuming their water content to be $\sim10\%$ by mass \citep{EOE2007,Sephton2002,Sephton2014}. Recent study by \citet{Rivkin2019} shows that C-type asteroids appear to have a significantly lower water content than chondrite meteorites. They argue that only a certain \revrevrev{fraction}, around 40\% of C type asteroids should be considered water rich. If the above-mentioned values have been adopted in this paper, it would bring the asteroid delivered water flux to Mercury down by 40\%. }

IDPs, comets, and C-type asteroids are not only rich in water but also in organic molecules.  While an endogenous contribution to Mercury's water cannot be ruled out, there is no plausible endogenous formation mechanism for organics; therefore any positive detection of organics would prove the exogenous origin of the bright and dark deposits \citep{Zhang2009}. 

For a better understanding of the nature and the origin of the dark and bright deposits in the polar regions of Mercury, more observational data are needed. 
\rev{For example, investigating various isotopic ratios, like D/H and $^{14}$N/$^{15}$N for the Earth, may be a way to constrain the origins of Mercury's water.} The joint ESA-JAXA mission BepiColombo, \rev{launched in 2018 and scheduled} to arrive at Mercury in 2025, will guide further exploration. \rev{BepiColombo's polar orbit will be much less eccentric than that of MESSENGER. This will allow to perform elemental measurements of the southern hemisphere that were made with poor spatial resolution or even not possible before.} Importantly for our purposes, \rev{BepiColombo} carries the Mercury Gamma-Ray and Neutron Spectrometer (MGNS), similar to MESSSENGER's NS but at higher resolution.  MGNS will map water across the entire surface of Mercury, including the Southern polar regions, down to a depth of 1--2\,m \citep{mitrofanov2010} and will clarify if the Southern polar \rev{region} is indeed as water-rich as its Northern counterpart.

\section{Conclusions}\label{conclusions}
IDPs, asteroids and comets play an important role in the formation process of the dark and bright deposits in the polar regions of Mercury that have been associated with water ice. \rev{While other sources are not ruled out by our analysis, we show that impacts can deliver a sufficient amount of water to Mercury's polar regions} to explain the available observational data\rev{; delivery would take \rev{$\sim 1$ Gyr}}.  \rev{IDPs deliver more water than asteroids and comets combined.}

\section{Acknowledgments}\label{aknowledgments}

This paper is dedicated to the brave resistance of Ukrainian people. \\[5pt]
\rev{We are thankful to \emph{David E. Kaufmann} for valuable help with RMVS/Swifter,} \emph{Cecile Engrand} and \emph{Mikhail Zelensky} for input on the water fraction of IDPs. This research has made use of data and/or services provided by the International Astronomical Union's Minor Planet Center. We would like to thank the Center for Information Technology of the University of Groningen for their support and for providing access to the Peregrine high performance computing cluster. Petr Pokorn\'{y}'s work was supported by NASA awards number 80GSFC21M0002 and 80NSSC21K0153. \rev{The authors thank the two anonymous reviewers for their thoughtful comments, which significantly improved the manuscript.}

\section*{References}

\bibliography{mybibfile}

\begin{thebibliography}{67}
\expandafter\ifx\csname natexlab\endcsname\relax\def\natexlab#1{#1}\fi
\expandafter\ifx\csname url\endcsname\relax
  \def\url#1{\texttt{#1}}\fi
\expandafter\ifx\csname urlprefix\endcsname\relax\def\urlprefix{URL }\fi

\bibitem[{{Borin} et~al.(2017){Borin}, {Cremonese}, {Marzari}, and
  {Lucchetti}}]{Borin2017}
{Borin}, P., {Cremonese}, G., {Marzari}, F., {Lucchetti}, A., 2017. {Asteroidal
  and cometary dust flux in the inner solar system}. \aap.

\bibitem[{{Bowell} et~al.(1989){Bowell}, {Hapke}, {Domingue}, {Lumme},
  {Peltoniemi}, and {Harris}}]{Bowell1989}
{Bowell}, E., {Hapke}, B., {Domingue}, D., {Lumme}, K., {Peltoniemi}, J.,
  {Harris}, A.~W., 1989. {Application of photometric models to asteroids}. In:
  {Binzel}, R.~P., {Gehrels}, T., {Matthews}, M.~S. (Eds.), Asteroids II. pp.
  524--556.

\bibitem[{{Butler}(1997)}]{Butler1997}
{Butler}, B.~J., Aug. 1997. {The migration of volatiles on the surfaces of
  Mercury and the Moon}. \jgr 102, 19283--19292.

\bibitem[{{Butler} et~al.(1993){Butler}, {Muhleman}, and {Slade}}]{Butler1993}
{Butler}, B.~J., {Muhleman}, D.~O., {Slade}, M.~A., Aug. 1993. {Mercury -
  Full-disk radar images and the detection and stability of ice at the North
  Pole}. \jgr 98, 15.

\bibitem[{{Carrillo-S{\'a}nchez} et~al.(2016){Carrillo-S{\'a}nchez},
  {Nesvorn{\'y}}, {Pokorn{\'y}}, {Janches}, and {Plane}}]{Carrillo-Sanchez2016}
{Carrillo-S{\'a}nchez}, J.~D., {Nesvorn{\'y}}, D., {Pokorn{\'y}}, P.,
  {Janches}, D., {Plane}, J.~M.~C., Dec. 2016. {Sources of cosmic dust in the
  Earth's atmosphere}. \grl 43, 11.

\bibitem[{{Carry}(2012)}]{Carry2012}
{Carry}, B., Dec. 2012. {Density of asteroids}. \planss 73, 98--118.

\bibitem[{{Ceplecha} et~al.(1998){Ceplecha}, {Borovi{\v{c}}ka}, {Elford},
  {Revelle}, {Hawkes}, {Porub{\v{c}}an}, and {{\v{S}}imek}}]{Ceplecha1998}
{Ceplecha}, Z., {Borovi{\v{c}}ka}, J., {Elford}, W.~G., {Revelle}, D.~O.,
  {Hawkes}, R.~L., {Porub{\v{c}}an}, V., {{\v{S}}imek}, M., Sep 1998. {Meteor
  Phenomena and Bodies}. \ssr 84, 327--471.

\bibitem[{{Ceplecha} and {Revelle}(2005)}]{Ceplecha2005}
{Ceplecha}, Z., {Revelle}, D.~O., Jan 2005. {Fragmentation model of meteoroid
  motion, mass loss, and radiation in the atmosphere}. Meteoritics and
  Planetary Science 40, 35.

\bibitem[{{Chabot} et~al.(2012){Chabot}, {Ernst}, {Denevi}, {Harmon},
  {Murchie}, {Blewett}, {Solomon}, and {Zhong}}]{chabot2012}
{Chabot}, N.~L., {Ernst}, C.~M., {Denevi}, B.~W., {Harmon}, J.~K., {Murchie},
  S.~L., {Blewett}, D.~T., {Solomon}, S.~C., {Zhong}, E.~D., May 2012. {Areas
  of permanent shadow in Mercury's south polar region ascertained by MESSENGER
  orbital imaging}. \grl 39, L09204.

\bibitem[{{Cintala}(1992)}]{Cintala1992}
{Cintala}, M.~J., Jan. 1992. {Impact-induced thermal effects in the lunar and
  mercurian regoliths}. \jgr 97~(E1), 947--973.

\bibitem[{{Cremonese} et~al.(2012){Cremonese}, {Borin}, {Martellato},
  {Marzari}, and {Bruno}}]{Cremonese2012}
{Cremonese}, G., {Borin}, P., {Martellato}, E., {Marzari}, F., {Bruno}, M., Apr
  2012. {New Calibration of the Micrometeoroid Flux on Earth}. \apjl 749~(2),
  L40.

\bibitem[{{Crider} and {Killen}(2005)}]{crider2005}
{Crider}, D., {Killen}, R.~M., Jun. 2005. {Burial rate of Mercury's polar
  volatile deposits}. \grl 32, L12201.

\bibitem[{{DeMeo} and {Carry}(2013)}]{DeMeo2013}
{DeMeo}, F.~E., {Carry}, B., Sep. 2013. {The taxonomic distribution of
  asteroids from multi-filter all-sky photometric surveys}. \icarus 226,
  723--741.

\bibitem[{{Denevi} et~al.(2013){Denevi}, {Ernst}, {Meyer}, {Robinson},
  {Murchie}, {Whitten}, {Head}, {Watters}, {Solomon}, {Ostrach}, {Chapman},
  {Byrne}, {Klimczak}, and {Peplowski}}]{Denevi2013}
{Denevi}, B.~W., {Ernst}, C.~M., {Meyer}, H.~M., {Robinson}, M.~S., {Murchie},
  S.~L., {Whitten}, J.~L., {Head}, J.~W., {Watters}, T.~R., {Solomon}, S.~C.,
  {Ostrach}, L.~R., {Chapman}, C.~R., {Byrne}, P.~K., {Klimczak}, C.,
  {Peplowski}, P.~N., May 2013. {The distribution and origin of smooth plains
  on Mercury}. Journal of Geophysical Research (Planets) 118~(5), 891--907.

\bibitem[{{Dobrica} et~al.(2010){Dobrica}, {Engrand}, {Duprat}, and
  {Gounelle}}]{Dobrica2010}
{Dobrica}, E., {Engrand}, C., {Duprat}, J., {Gounelle}, M., Sep. 2010. {A
  Statistical Overview of CONCORDIA Antarctic Micrometeorites}. Meteoritics and
  Planetary Science Supplement 73, 5213.

\bibitem[{{Eke} et~al.(2017){Eke}, {Lawrence}, and {Teodoro}}]{Eke2017}
{Eke}, V.~R., {Lawrence}, D.~J., {Teodoro}, L.~F.~A., Mar. 2017. {How thick are
  Mercury's polar water ice deposits?} \icarus 284, 407--415.

\bibitem[{{Engrand} et~al.(1996){Engrand}, {Deloule}, {Hoppe}, {Kurat},
  {Maurette}, and {Robert}}]{Engrand1996}
{Engrand}, C., {Deloule}, E., {Hoppe}, P., {Kurat}, G., {Maurette}, M.,
  {Robert}, F., Mar. 1996. {Water Contents of Micrometeorites from Antarctica}.
  In: Lunar and Planetary Science Conference. Vol.~27 of Lunar and Planetary
  Science Conference.

\bibitem[{{Engrand} et~al.(1999){Engrand}, {Deloule}, {Robert}, {Maurette}, and
  {Kurat}}]{Engrand1999}
{Engrand}, C., {Deloule}, E., {Robert}, F., {Maurette}, M., {Kurat}, G., Sep.
  1999. {Extraterrestrial water in micrometeorites and cosmic spherules from
  Antarctica: an ion microprobe study}. Meteoritics and Planetary Science 34,
  773--786.

\bibitem[{{Fassett} et~al.(2012){Fassett}, {Head}, {Baker}, {Zuber}, {Smith},
  {Neumann}, {Solomon}, {Klimczak}, {Strom}, {Chapman}, {Prockter}, {Phillips},
  {Oberst}, and {Preusker}}]{Fassett2012}
{Fassett}, C.~I., {Head}, J.~W., {Baker}, D.~M.~H., {Zuber}, M.~T., {Smith},
  D.~E., {Neumann}, G.~A., {Solomon}, S.~C., {Klimczak}, C., {Strom}, R.~G.,
  {Chapman}, C.~R., {Prockter}, L.~M., {Phillips}, R.~J., {Oberst}, J.,
  {Preusker}, F., Oct. 2012. {Large impact basins on Mercury: Global
  distribution, characteristics, and modification history from MESSENGER
  orbital data}. Journal of Geophysical Research (Planets) 117, E00L08.

\bibitem[{{Frantseva} et~al.(2018){Frantseva}, {Mueller}, {ten Kate}, {van der
  Tak}, and {Greenstreet}}]{Frantseva2018}
{Frantseva}, K., {Mueller}, M., {ten Kate}, I.~L., {van der Tak}, F.~F.~S.,
  {Greenstreet}, S., Jul. 2018. {Delivery of organics to Mars through asteroid
  and comet impacts}. \icarus 309, 125--133.

\bibitem[{{Gicquel} et~al.(2012){Gicquel}, {Bockel{\'e}e-Morvan}, {Zakharov},
  {Kelley}, {Woodward}, and {Wooden}}]{Gicquel2012}
{Gicquel}, A., {Bockel{\'e}e-Morvan}, D., {Zakharov}, V.~V., {Kelley}, M.~S.,
  {Woodward}, C.~E., {Wooden}, D.~H., Jun. 2012. {Investigation of dust and
  water ice in comet 9P/Tempel 1 from Spitzer observations of the Deep Impact
  event}. \aap 542, A119.

\bibitem[{{Grun} et~al.(1985){Grun}, {Zook}, {Fechtig}, and {Giese}}]{Grun1985}
{Grun}, E., {Zook}, H.~A., {Fechtig}, H., {Giese}, R.~H., May 1985.
  {Collisional balance of the meteoritic complex}. \icarus 62, 244--272.

\bibitem[{{Harmon} and {Slade}(1992)}]{Harmon1992}
{Harmon}, J.~K., {Slade}, M.~A., Oct. 1992. {Radar mapping of Mercury -
  Full-disk images and polar anomalies}. Science 258, 640--643.

\bibitem[{{Harmon} et~al.(1994){Harmon}, {Slade}, {V{\'e}lez}, {Crespo},
  {Dryer}, and {Johnson}}]{Harmon1994}
{Harmon}, J.~K., {Slade}, M.~A., {V{\'e}lez}, R.~A., {Crespo}, A., {Dryer},
  M.~J., {Johnson}, J.~M., May 1994. {Radar mapping of Mercury's polar
  anomalies}. \nat 369, 213--215.

\bibitem[{{Head} et~al.(2009){Head}, {Murchie}, {Prockter}, {Solomon},
  {Chapman}, {Strom}, {Watters}, {Blewett}, {Gillis-Davis}, {Fassett},
  {Dickson}, {Morgan}, and {Kerber}}]{Head2009}
{Head}, J.~W., {Murchie}, S.~L., {Prockter}, L.~M., {Solomon}, S.~C.,
  {Chapman}, C.~R., {Strom}, R.~G., {Watters}, T.~R., {Blewett}, D.~T.,
  {Gillis-Davis}, J.~J., {Fassett}, C.~I., {Dickson}, J.~L., {Morgan}, G.~A.,
  {Kerber}, L., Aug 2009. {Volcanism on Mercury: Evidence from the first
  MESSENGER flyby for extrusive and explosive activity and the volcanic origin
  of plains}. Earth and Planetary Science Letters 285~(3-4), 227--242.

\bibitem[{Hendrix et~al.(2019)Hendrix, Hurley, Farrell, Greenhagen, Hayne,
  Retherford, Vilas, Cahill, Poston, and Liu}]{Hendrix2019}
Hendrix, A.~R., Hurley, D.~M., Farrell, W.~M., Greenhagen, B.~T., Hayne, P.~O.,
  Retherford, K.~D., Vilas, F., Cahill, J. T.~S., Poston, M.~J., Liu, Y., 2019.
  Diurnally-migrating lunar water: Evidence from ultraviolet data. Geophysical
  Research Letters 46, 2417--2424.
\newline\urlprefix\url{https://agupubs.onlinelibrary.wiley.com/doi/abs/10.1029/2018GL081821}

\bibitem[{{Huebner}(2002)}]{Huebner2002}
{Huebner}, W.~F., Oct. 2002. {Composition of Comets: Observations and Models}.
  Earth Moon and Planets 89, 179--195.

\bibitem[{{Jewitt}(2004)}]{Jewitt2004}
{Jewitt}, D.~C., 2004. {From cradle to grave: the rise and demise of the
  comets}. pp. 659--676.

\bibitem[{{Jutzi} et~al.(2015){Jutzi}, {Holsapple}, {W{\"u}nneman}, and
  {Michel}}]{Jutzi2015}
{Jutzi}, M., {Holsapple}, K., {W{\"u}nneman}, K., {Michel}, P., Feb. 2015.
  {Modeling asteroid collisions and impact processes}. ArXiv e-prints.

\bibitem[{{Kessler}(1981)}]{Kessler1981}
{Kessler}, D.~J., Oct. 1981. {Derivation of the collision probability between
  orbiting objects: the lifetimes of jupiter's outer moons}. \icarus 48~(1),
  39--48.

\bibitem[{{Lawrence} et~al.(2013){Lawrence}, {Feldman}, {Goldsten}, {Maurice},
  {Peplowski}, {Anderson}, {Bazell}, {McNutt}, {Nittler}, {Prettyman},
  {Rodgers}, {Solomon}, and {Weider}}]{Lawrence2013}
{Lawrence}, D.~J., {Feldman}, W.~C., {Goldsten}, J.~O., {Maurice}, S.,
  {Peplowski}, P.~N., {Anderson}, B.~J., {Bazell}, D., {McNutt}, R.~L.,
  {Nittler}, L.~R., {Prettyman}, T.~H., {Rodgers}, D.~J., {Solomon}, S.~C.,
  {Weider}, S.~Z., Jan. 2013. {Evidence for Water Ice Near Mercury's North Pole
  from MESSENGER Neutron Spectrometer Measurements}. Science 339, 292.

\bibitem[{{Levison} and {Duncan}(1994)}]{Levison1994}
{Levison}, H.~F., {Duncan}, M.~J., Mar. 1994. {The long-term dynamical behavior
  of short-period comets}. \icarus 108, 18--36.

\bibitem[{{Love} and {Brownlee}(1993)}]{Love1993}
{Love}, S.~G., {Brownlee}, D.~E., Oct. 1993. {A Direct Measurement of the
  Terrestrial Mass Accretion Rate of Cosmic Dust}. Science 262, 550--553.

\bibitem[{{Mitrofanov} et~al.(2010){Mitrofanov}, {Kozyrev}, {Konovalov},
  {Litvak}, {Malakhov}, {Mokrousov}, {Sanin}, {Tret'ykov}, {Vostrukhin},
  {Bobrovnitskij}, {Tomilina}, {Gurvits}, and {Owens}}]{mitrofanov2010}
{Mitrofanov}, I.~G., {Kozyrev}, A.~S., {Konovalov}, A., {Litvak}, M.~L.,
  {Malakhov}, A.~A., {Mokrousov}, M.~I., {Sanin}, A.~B., {Tret'ykov}, V.~I.,
  {Vostrukhin}, A.~V., {Bobrovnitskij}, Y.~I., {Tomilina}, T.~M., {Gurvits},
  L., {Owens}, A., Jan. 2010. {The Mercury Gamma and Neutron Spectrometer
  (MGNS) on board the Planetary Orbiter of the BepiColombo mission}. \planss
  58, 116--124.

\bibitem[{{Morbidelli} et~al.(2018){Morbidelli}, {Nesvorny}, {Laurenz},
  {Marchi}, {Rubie}, {Elkins-Tanton}, {Wieczorek}, and
  {Jacobson}}]{Morbidelli2018}
{Morbidelli}, A., {Nesvorny}, D., {Laurenz}, V., {Marchi}, S., {Rubie}, D.~C.,
  {Elkins-Tanton}, L., {Wieczorek}, M., {Jacobson}, S., May 2018. {The timeline
  of the lunar bombardment: Revisited}. \icarus 305, 262--276.

\bibitem[{{Moses} et~al.(1999){Moses}, {Rawlins}, {Zahnle}, and
  {Dones}}]{Moses1999}
{Moses}, J.~I., {Rawlins}, K., {Zahnle}, K., {Dones}, L., Feb. 1999. {External
  Sources of Water for Mercury's Putative Ice Deposits}. \icarus 137, 197--221.

\bibitem[{{National Research Council}(2007)}]{EOE2007}
{National Research Council}, 2007. {Exploring Organic Environments in the Solar
  System}.

\bibitem[{{Nesvorn{\'y}} et~al.(2011{\natexlab{a}}){Nesvorn{\'y}}, {Janches},
  {Vokrouhlick{\'y}}, {Pokorn{\'y}}, {Bottke}, and
  {Jenniskens}}]{Nesvorny2011a}
{Nesvorn{\'y}}, D., {Janches}, D., {Vokrouhlick{\'y}}, D., {Pokorn{\'y}}, P.,
  {Bottke}, W.~F., {Jenniskens}, P., Dec. 2011{\natexlab{a}}. {Dynamical Model
  for the Zodiacal Cloud and Sporadic Meteors}. \apj 743, 129.

\bibitem[{{Nesvorn{\'y}} et~al.(2010){Nesvorn{\'y}}, {Jenniskens}, {Levison},
  {Bottke}, {Vokrouhlick{\'y}}, and {Gounelle}}]{Nesvorny2010}
{Nesvorn{\'y}}, D., {Jenniskens}, P., {Levison}, H.~F., {Bottke}, W.~F.,
  {Vokrouhlick{\'y}}, D., {Gounelle}, M., Apr. 2010. {Cometary Origin of the
  Zodiacal Cloud and Carbonaceous Micrometeorites. Implications for Hot Debris
  Disks}. \apj 713, 816--836.

\bibitem[{{Nesvorn{\'y}} et~al.(2017){Nesvorn{\'y}}, {Vokrouhlick{\'y}},
  {Dones}, {Levison}, {Kaib}, and {Morbidelli}}]{Nesvorny2017}
{Nesvorn{\'y}}, D., {Vokrouhlick{\'y}}, D., {Dones}, L., {Levison}, H.~F.,
  {Kaib}, N., {Morbidelli}, A., Aug. 2017. {Origin and Evolution of
  Short-period Comets}. \apj 845~(1), 27.

\bibitem[{{Nesvorn{\'y}} et~al.(2011{\natexlab{b}}){Nesvorn{\'y}},
  {Vokrouhlick{\'y}}, {Pokorn{\'y}}, and {Janches}}]{Nesvorny2011b}
{Nesvorn{\'y}}, D., {Vokrouhlick{\'y}}, D., {Pokorn{\'y}}, P., {Janches}, D.,
  Dec. 2011{\natexlab{b}}. {Dynamics of Dust Particles Released from Oort Cloud
  Comets and Their Contribution to Radar Meteors}. \apj 743, 37.

\bibitem[{{Neumann} et~al.(2013){Neumann}, {Cavanaugh}, {Sun}, {Mazarico},
  {Smith}, {Zuber}, {Mao}, {Paige}, {Solomon}, {Ernst}, and
  {Barnouin}}]{Neumann2013}
{Neumann}, G.~A., {Cavanaugh}, J.~F., {Sun}, X., {Mazarico}, E.~M., {Smith},
  D.~E., {Zuber}, M.~T., {Mao}, D., {Paige}, D.~A., {Solomon}, S.~C., {Ernst},
  C.~M., {Barnouin}, O.~S., Jan. 2013. {Bright and Dark Polar Deposits on
  Mercury: Evidence for Surface Volatiles}. Science 339, 296.

\bibitem[{{Nittler} et~al.(2017){Nittler}, {Chabot}, {Grove}, and
  {Peplowski}}]{Nittler2017}
{Nittler}, L.~R., {Chabot}, N.~L., {Grove}, T.~L., {Peplowski}, P.~N., Dec
  2017. {The Chemical Composition of Mercury}. arXiv e-prints,
  arXiv:1712.02187.

\bibitem[{{Noguchi} et~al.(2002){Noguchi}, {Nakamura}, and
  {Nozaki}}]{Noguchi2002}
{Noguchi}, T., {Nakamura}, T., {Nozaki}, W., Sep. 2002. {Mineralogy of
  phyllosilicate-rich micrometeorites and comparison with Tagish Lake and
  Sayama meteorites}. Earth and Planetary Science Letters 202, 229--246.

\bibitem[{{Ostrach} et~al.(2015){Ostrach}, {Robinson}, {Whitten}, {Fassett},
  {Strom}, {Head}, and {Solomon}}]{Ostrach2015}
{Ostrach}, L.~R., {Robinson}, M.~S., {Whitten}, J.~L., {Fassett}, C.~I.,
  {Strom}, R.~G., {Head}, J.~W., {Solomon}, S.~C., Apr 2015. {Extent, age, and
  resurfacing history of the northern smooth plains on Mercury from MESSENGER
  observations}. \icarus 250, 602--622.

\bibitem[{{Paige} et~al.(2013){Paige}, {Siegler}, {Harmon}, {Neumann},
  {Mazarico}, {Smith}, {Zuber}, {Harju}, {Delitsky}, and {Solomon}}]{Paige2013}
{Paige}, D.~A., {Siegler}, M.~A., {Harmon}, J.~K., {Neumann}, G.~A.,
  {Mazarico}, E.~M., {Smith}, D.~E., {Zuber}, M.~T., {Harju}, E., {Delitsky},
  M.~L., {Solomon}, S.~C., Jan. 2013. {Thermal Stability of Volatiles in the
  North Polar Region of Mercury}. Science 339, 300.

\bibitem[{{Pokorn{\'y}} et~al.(2017){Pokorn{\'y}}, {Sarantos}, and
  {Janches}}]{Pokorny2017}
{Pokorn{\'y}}, P., {Sarantos}, M., {Janches}, D., Jun. 2017. {Reconciling the
  Dawn-Dusk Asymmetry in Mercury's Exosphere with the Micrometeoroid Impact
  Directionality}. \apjl 842, L17.

\bibitem[{{Pokorn{\'y}} et~al.(2018){Pokorn{\'y}}, {Sarantos}, and
  {Janches}}]{Pokorny2018}
{Pokorn{\'y}}, P., {Sarantos}, M., {Janches}, D., Aug. 2018. {A Comprehensive
  Model of the Meteoroid Environment around Mercury}. \apj 863, 31.

\bibitem[{{Pokorn{\'y}} and {Vokrouhlick{\'y}}(2013)}]{Pokorny2013}
{Pokorn{\'y}}, P., {Vokrouhlick{\'y}}, D., Sep. 2013. {{\"O}pik-type collision
  probability for high-inclination orbits: Targets on eccentric orbits}.
  \icarus 226~(1), 682--693.

\bibitem[{{Pokorn{\'y}} et~al.(2014){Pokorn{\'y}}, {Vokrouhlick{\'y}},
  {Nesvorn{\'y}}, {Campbell-Brown}, and {Brown}}]{Pokorny2014}
{Pokorn{\'y}}, P., {Vokrouhlick{\'y}}, D., {Nesvorn{\'y}}, D.,
  {Campbell-Brown}, M., {Brown}, P., Jul. 2014. {Dynamical Model for the
  Toroidal Sporadic Meteors}. \apj 789~(1), 25.

\bibitem[{{Potter}(1995)}]{Potter1995}
{Potter}, A.~E., 1995. {Chemical sputtering could produce sodium vapor and ice
  on Mercury}. \grl 22, 3289--3292.

\bibitem[{{Pravec} and {Harris}(2007)}]{Pravec2007}
{Pravec}, P., {Harris}, A.~W., Sep. 2007. {Binary asteroid population. 1.
  Angular momentum content}. \icarus 190, 250--259.

\bibitem[{{Prialnik}(2002)}]{Prialnik2002}
{Prialnik}, D., Oct. 2002. {Modeling the Comet Nucleus Interior}. Earth Moon
  and Planets 89, 27--52.

\bibitem[{{Prockter} et~al.(2010){Prockter}, {Ernst}, {Denevi}, {Chapman},
  {Head}, {Fassett}, {Merline}, {Solomon}, {Watters}, {Strom}, {Cremonese},
  {Marchi}, and {Massironi}}]{Prockter2010}
{Prockter}, L.~M., {Ernst}, C.~M., {Denevi}, B.~W., {Chapman}, C.~R., {Head},
  J.~W., {Fassett}, C.~I., {Merline}, W.~J., {Solomon}, S.~C., {Watters},
  T.~R., {Strom}, R.~G., {Cremonese}, G., {Marchi}, S., {Massironi}, M., Aug
  2010. {Evidence for Young Volcanism on Mercury from the Third MESSENGER
  Flyby}. Science 329~(5992), 668.

\bibitem[{{Rivkin} and {DeMeo}(2019)}]{Rivkin2019}
{Rivkin}, A.~S., {DeMeo}, F.~E., Jan. 2019. {How Many Hydrated NEOs Are There?}
  Journal of Geophysical Research (Planets) 124~(1), 128--142.

\bibitem[{{Sephton}(2014)}]{Sephton2014}
{Sephton}, M.~A., Oct. 2014. {Organic Geochemistry of Meteorites}.

\bibitem[{{Sephton} et~al.(2002){Sephton}, {Wright}, {Gilmour}, {de Leeuw},
  {Grady}, and {Pillinger}}]{Sephton2002}
{Sephton}, M.~A., {Wright}, I.~P., {Gilmour}, I., {de Leeuw}, J.~W., {Grady},
  M.~M., {Pillinger}, C.~T., Jun. 2002. {High molecular weight organic matter
  in martian meteorites}. Planetary and Space Sciences Research Institute 50,
  711--716.

\bibitem[{{Slade} et~al.(1992){Slade}, {Butler}, and {Muhleman}}]{Slade1992}
{Slade}, M.~A., {Butler}, B.~J., {Muhleman}, D.~O., Oct. 1992. {Mercury radar
  imaging - Evidence for polar ice}. Science 258, 635--640.

\bibitem[{{Svetsov}(2011)}]{Svetsov2011}
{Svetsov}, V., Jul. 2011. {Cratering erosion of planetary embryos}. \icarus
  214, 316--326.

\bibitem[{{Swamy}(2010)}]{Swamy2010}
{Swamy}, K.~S.~K., 2010. {Physics of Comets (3rd Edition)}. World Scientific
  Publishing Co.

\bibitem[{{Taylor} et~al.(2017){Taylor}, {Altobelli}, {Buratti}, and
  {Choukroun}}]{Taylor2017}
{Taylor}, M.~G.~G.~T., {Altobelli}, N., {Buratti}, B.~J., {Choukroun}, M., May
  2017. {The Rosetta mission orbiter science overview: the comet phase}.
  Philosophical Transactions of the Royal Society of London Series A 375,
  20160262.

\bibitem[{{Vokrouhlick{\'y}} et~al.(2019){Vokrouhlick{\'y}}, {Nesvorn{\'y}},
  and {Dones}}]{Vokrouhlicky2019}
{Vokrouhlick{\'y}}, D., {Nesvorn{\'y}}, D., {Dones}, L., May 2019. {Origin and
  Evolution of Long-period Comets}. \aj 157~(5), 181.

\bibitem[{{Wiegert} et~al.(2009){Wiegert}, {Vaubaillon}, and
  {Campbell-Brown}}]{Wiegert2009}
{Wiegert}, P., {Vaubaillon}, J., {Campbell-Brown}, M., May 2009. {A dynamical
  model of the sporadic meteoroid complex}. \icarus 201~(1), 295--310.

\bibitem[{{Wilson} and {Head}(2008)}]{Wilson2008}
{Wilson}, L., {Head}, J.~W., Dec 2008. {Volcanism on Mercury: A new model for
  the history of magma ascent and eruption}. \grl 35~(23), L23205.

\bibitem[{{Wright} et~al.(2016){Wright}, {Mainzer}, {Masiero}, {Grav}, and
  {Bauer}}]{Wright2016}
{Wright}, E.~L., {Mainzer}, A., {Masiero}, J., {Grav}, T., {Bauer}, J., Oct.
  2016. {The Albedo Distribution of Near Earth Asteroids}. \aj 152, 79.

\bibitem[{{Zhang} and {Paige}(2009)}]{Zhang2009}
{Zhang}, J.~A., {Paige}, D.~A., Aug. 2009. {Cold-trapped organic compounds at
  the poles of the Moon and Mercury: Implications for origins}. \grl 36,
  L16203.

\bibitem[{{Zolensky} and {Lindstrom}(1992)}]{Zolensky1992}
{Zolensky}, M.~E., {Lindstrom}, D.~J., 1992. {Mineralogy of 12 large
  'chondritic' interplanetary dust particles}. In: {Ryder}, G., {Sharpton},
  V.~L. (Eds.), Lunar and Planetary Science Conference Proceedings. Vol.~22 of
  Lunar and Planetary Science Conference Proceedings. pp. 161--169.

\end{thebibliography}

\end{document}